\documentclass{article}

\usepackage{arxiv}

\usepackage[utf8]{inputenc} 
\usepackage[T1]{fontenc}    
\usepackage{url}            
\usepackage{booktabs}       
\usepackage{amsfonts}       
\usepackage{nicefrac}       
\usepackage{microtype}      
\usepackage{lipsum}

\usepackage{amssymb}
\usepackage{color, soul}
\usepackage{amsmath}
\usepackage{tabularx}
\usepackage{caption}
\usepackage{booktabs}
\usepackage{graphicx}
\usepackage{multirow}
\usepackage[hidelinks]{hyperref}

\title{A latent variable approach to account for correlated inputs in global sensitivity analysis with cases from pharmacological systems modelling}

\author{
  Nicola Melillo \\
    Centre for Applied Pharmacokinetic Research \\
  School of Health Sciences\\
  The University of Manchester\\
  Manchester, UK \\
  \texttt{nicola.melillo@manchester.ac.uk} \\
   \And
 Adam S. Darwich \\
  Division of Health Informatics and Logistics\\
  Department of Biomedical Engineering and Health Systems \\
  KTH Royal Institute of Technology\\
  Stockholm, Sweden \\
  \texttt{darwich@kth.se} \\
}

\begin{document}
\maketitle

\begin{abstract}
In pharmaceutical research and development decision-making related to drug candidate selection, efficacy and safety is commonly supported through modelling and simulation (M\&S). Among others, physiologically-based pharmacokinetic models are used to describe drug absorption, distribution and metabolism in human. Global sensitivity analysis (GSA) is gaining interest in the pharmacological M\&S community as an important element for quality assessment of model-based inference. Physiological models often present inter-correlated parameters. The inclusion of correlated factors in GSA and the sensitivity indices interpretation has proven an issue for these models.
Here we devise and evaluate a latent variable approach for dealing with correlated factors in GSA. This approach describes the correlation between two model inputs through the causal relationship of three independent factors: the latent variable and the unique variances of the two correlated parameters. Then, GSA is performed with the classical variance-based method. We applied the latent variable approach to a set of algebraic models and a case from physiologically-based pharmacokinetics. Then, we compared our approach to Sobol's GSA assuming no correlations, Sobol's GSA with groups and the Kucherenko approach.
The relative ease of implementation and interpretation makes this a simple approach for carrying out GSA for models with correlated input factors.
\end{abstract}



\clearpage
\section{Introduction}
\label{section_intro}

The use of sensitivity analysis (SA), including global SA (GSA), has gained interest from pharmaceutical industry, regulators and academia in recent years \cite{chmp_ema_guideline_2018,fda_pbpkguide_2018,melillo_variance_2019,melillo_accounting_2019,melillo_inter-compound_2020,mcnally_workflow_2011,zhang_sobol_2015,daga_physiologically_2018-1,yau_gsappkrr_2020,liu_investigating_2019}. In pharmaceutical research and development (R\&D) decision-making for drug candidate selection, efficacy and safety is often supported by modelling and simulation (M\&S). This is referred to as \textbf{m}odel-\textbf{i}nformed \textbf{d}rug \textbf{d}iscovery and \textbf{d}evelopment (MID3) \cite{efpia_mid3goodpractices_2016}. SA is an indispensable instrument for the quality assessment of model-based inference \cite{saltelli_what_2013}. However, it is the authors' opinion, that the treatment and interpretation of correlated input parameters in GSA can be a barrier to wider use.

\subsection{Modelling for decision-making in pharmaceutical R\&D}
Broadly, modelling activities in pharmaceutical R\&D are centred around the study of disease, pharmacokinetics (PK; \textit{in vivo} drug \textbf{a}bsorption, \textbf{d}istribution, \textbf{m}etabolism and \textbf{e}limination, or ADME), structure-activity relationships, pharmacodynamics (PD; temporal pharmacological effects) and more \cite{efpia_mid3goodpractices_2016}. In this work, we focus on PK models.

PK models vary in complexity, ranging from simple empirical models, to complex models based on physiological considerations \cite{aarons_physiologically_2005, bonate_poppkpd_2005}. Empirical PK models use functions, such as sum of exponential terms, to describe the drug concentration-time profiles. In these models the parameters are generally estimated from the data and have no clear physiological meaning. Here we focus on the class of physiologically-based models, such as systems models of biology, disease and pharmacology, that are applied throughout drug development. As these models are based on physiological mechanisms, they can be used to extrapolate PK/PD effects from \textit{in vitro} to \textit{in vivo}, between species, populations and scenarios \cite{jones_basic_2013}. 

Physiologically-based pharmacokinetic (PBPK) M\&S provides a framework for mechanistic predictions of PK. PBPK models consist of systems of ordinary differential equations based on mass balance (see also physiologically-based toxicokinetics \cite{sasso_pbtk_2010}). PBPK models are compartmental models in which each compartment corresponds to a specific organ or tissue and is connected to other compartments through flow rates representing the blood circulation. This structure reflects a representation of the true anatomical layout. Data on population demographics, tissue composition, organ function, drug metabolising enzymes, transporters and whole blood parameters are integrated with drug and formulation-specific information to predict drug exposure over time \cite{nestorov_wbpbpk_2007}. PBPK M\&S has been used to replace/supplement clinical trials and inform labelling for numerous drugs, most notably for dosage recommendations following metabolic drug-drug interactions \cite{jamei_pbpklabel_2016,grimstein_physiologically_2019}, avoiding adverse effects in patients.

Uncertainty and variability are prominent in biological data and an important consideration for decisions on drug safety. For example, during drug development, when the drug is administered for the first time in humans (so called, first-in-human trials), uncertainty affects the probability in risk predictions, therefore informing the dosing strategy. In this context, uncertainty mainly relates to inter- and intra-experimental variability, experimental errors in laboratory and physiological measures, and translation of parameters. Variability mainly relates to interindividual variability in physiology, protein expression, genetics, interoccasion variability and more. 

Correlations between input parameters are often implemented in PBPK models to account for physiological constraints, otherwise causing implausible combinations of parameters \cite{tsamandouras_incorporation_2015,rostamihodjegan_reverse_2018}. For example, organ weights and blood flows are constrained by body weight and cardiac output. With the emergence of novel `omics techniques, the correlation of proteins is also of increasing interest \cite{bai_omicspharmaceutics_2018,doki_pbpkcorr_2018}.

\subsection{Global sensitivity analysis for PBPK M\&S: the issue of correlated input factors}
Both the United States Food and Drug Administration (FDA) and European Medicines Agency (EMA) have highlighted the importance of SA and GSA as best practice in PBPK to inform model development and refinement \cite{chmp_ema_guideline_2018,fda_pbpkguide_2018}. GSA is key for elucidating the relationship between the uncertainty and variability in model inputs and variation in a given model output. Therefore, by extension, the method is also relevant for drug development and precision dosing in clinical practice \cite{melillo_variance_2019,melillo_accounting_2019,yau_gsappkrr_2020,garciacremades_mechanistic_2020,rowland_pbpkdrivers_2018}. 

In this work, we focused on the variance-based GSA (also referred to as Sobol's method) \cite{melillo_variance_2019,melillo_inter-compound_2020}. This choice was made as the variance-based GSA is able to handle nonlinear and nonmonotonic relationships between the input factors and the model outputs \cite{sobol_sensitivity_1993,saltelli_making_2002,saltelli_global_2008}. Moreover, with this method it is possible to quantify the effect of each factor taken singularly and the extent of its interaction effects. As we have reported in previous work, understanding the extent of the interaction effects can be particularly important for an informed use of PBPK models during drug development \cite{melillo_inter-compound_2020}.

The classical variance-based GSA works under the assumption that model inputs (commonly referred to as model parameters in pharmacometrics) are independent \cite{sobol_sensitivity_1993,saltelli_global_2008,saltelli_making_2002}. 
Under this assumption, the variance decomposition is unique \cite{sobol_sensitivity_1993} and reflects the structure of the model itself \cite{oakley_probabilistic_2004}. In this context, the variance-based sensitivity indices have a clear interpretation \cite{saltelli_making_2002,iooss_shapley_2019}. However, most PBPK models violate the independence assumption \cite{melillo_accounting_2019,tsamandouras_incorporation_2015,liu_considerations_2020}. In practice this may lead to correlations being ignored in the analysis, or the use of one of several proposed methods for GSA that deal with dependent inputs. Perhaps, the most simple and elegant way of treating dependent inputs in GSA is by grouping the correlated factors and then performing a GSA with the independent groups. The intrinsic limitation of this approach is that it is not possible to distinguish the contribution of the single variables within each group.

In the literature, several methods have been developed to deal with dependent inputs while retaining the information, or sensitivity indices, of each individual factor.
These methods typically fall into one of two classes: parametric and non-parametric methods \cite{mara_nonparam_2015,do_correlation_2020}. The parametric methods, also called model-based methods, (e.g., \cite{da_veiga_local_2009,li_global_2010,xu_extending_2007}) assume an \textit{a priori} model for the input-output relation. Instead, the non-parametric approaches do not assume any specific shape for this relation and thus, they are referred to as model-free or non model-based methods \cite{do_correlation_2020,mara_nonparam_2015}. These approaches are by and large considered more suitable for computer-based modelling \cite{do_correlation_2020}. Generally, the non-parametric methods employ a transformation technique for dealing with correlated factors' distribution \cite{do_correlation_2020}. For example, Kucherenko \textit{et al.} \cite{kucherenko_estimation_2012} used copula transformations to generalise the first order and total Sobol indices for the case of dependent input factors. Mara \textit{et al.} \cite{mara_nonparam_2015} proposed the use of the Rosenblatt transformation, and Tarantola and Mara \cite{tarantola_variance-based_2017} used both the Rosenblatt and Nataf transformation within the context of variance-based GSA. Moreover, other methods such as the variogram analysis of response surfaces (VARS) and the Shapely effects have been extended for the case of correlated input factors \cite{do_correlation_2020,iooss_shapley_2019}. 

The copula-based method, developed by Kucherenko and coworkers \cite{kucherenko_estimation_2012}, has recently been proposed for PBPK models and implemented in a commercial PBPK software \cite{liu_considerations_2020,noauthor_simcyp_2019}. However, how to interpret variance-based GSA results in presence of dependent variables is not straightforward and still debated among GSA practitioners. In presence of correlation between the input factors, the correspondence between the variance-based indices and model structure is lost and the variance decomposition can no longer provide a description of the model structure \cite{oakley_probabilistic_2004,borgonovo_sensitivity_2016,pianosi_sensitivity_2016}. This was illustrated by Oakley and O'Hagan in 2004 with the use of a simple example \cite{oakley_probabilistic_2004}. In this context, Pianosi \textit{et al.} reported that `\textit{counterintuitive results may be obtained}' \cite{pianosi_sensitivity_2016}. Iooss and Lema\^{\i}tre reported that `\textit{SA for dependent inputs has also been discussed by several authors [...], but this issue remains misunderstood}' \cite{iooss_review_2015}. Moreover, Iooss and Prieur reported that `\textit{The so-called Sobol' indices [...], present a difficult interpretation in the presence of statistical dependence between inputs}' \cite{iooss_shapley_2019}.

Several dedicated software platforms exist for PBPK M\&S \cite{kostewicz_pbpksoftware_2014}, providing accessible tools for non-expert users. As GSA gains use in the community (such as through software implementation) the issue of interpretability becomes increasingly relevant.


Here we propose a latent variable approach for treating correlated input parameters in variance-based GSA. The method expresses the correlation between two parameters as causal relationships between uncorrelated variables. This is done in order to allow the use of classical variance-based GSA and avoids the usage of methods whose interpretation is still a matter of debate. Latent variable models and sub-varieties of them, such as factor analysis, path analysis and structural equation modelling, are widely used in social sciences \cite{loehlin_latent_2017}. In latent variable models, the correlation between more than one observed measure (or model parameter) is described by one, or more, unobserved (latent) variable(s). Parameters are correlated as they share a common cause \cite{brown_confirmatory_2015}. Here we focus on the case of two linearly correlated random variables whose correlation is explained by one latent variable. With this approach, instead of two correlated factors, three independent factors (the latent variable and the two independent variances of the correlated parameters) are considered in the GSA.

The approach is then applied to a set of algebraic models and a whole-body PBPK model for the drug midazolam (MDZ). MDZ is a sedative primarily metabolised by Cythochrome P450 (CYP) 3A4 and CYP3A5 \cite{Galetin_midazolam_2004}. The expression of CYP3A5 is polymorphic and present in around 10-20\% \cite{roy_cyp3a5polymorphism_2005} of Caucasians where it is correlated with CYP3A4 through a shared mechanism for expression \cite{lolodi_pxrcyp3a_2017}. The latent variable approach was then compared with the classic Sobol's variance-based GSA, Sobol's GSA performed by grouping together the correlated factors, and the Kucherenko approach.
\clearpage
\section{Materials and Methods}
\label{section_methods}

\subsection{Variance-based sensitivity analysis and the Kucherenko approach}
Let us consider the generic model in Equation \ref{eq_generic_model}:
\begin{equation}\label{eq_generic_model}
    Y = f(\mathbf{X}),
\end{equation}
where $Y$ is the scalar model output, $\mathbf{X}$ is the vector including the $k$ independent input factors ($X_i$, $i=1...k$) and $f$ is the input-output relationship. In variance-based GSA (also known as Sobol's GSA) two sensitivity indices are derived from the decomposition of the variance ($V$) of $Y$. These are the so called first order index (or main effect) $S_i$ and the total effect ($S_{T,i}$), in Equation system \ref{eq_vb_indices} \cite{sobol_sensitivity_1993,saltelli_global_2008,homma_importance_1996}.
\begin{equation}\label{eq_vb_indices}
\begin{aligned}
S_{i} &= \dfrac{V_{X_i}(E_{\mathbf{X_{\sim i}}} (Y \, | \, X_i))}{V(Y)} \\
S_{Ti} &= \frac{E_{\mathbf{X}_{\sim i}} ( V_{X_i}( Y \, | \, \mathbf{X}_{\sim i} ) )}{V(Y)}
\end{aligned}
\end{equation}
$\mathbf{X}_{\sim i}$ represents a vector including all the factors except $X_i$, while $E$ is the expectation operator. $S_i$ is related with the part of $V(Y)$ explained by the variation of $X_i$ taken singularly and $S_{T,i}$ is the sum of $S_i$ with all the interaction effects of $X_i$ with the other inputs \cite{saltelli_making_2002,saltelli_global_2008}. When the parameters are independent, the relationships  $S_i \leq S_{T,i}$ and $\sum S_i \leq 1$ are always valid and $S_{T,i}-S_i$ gives information about the extent of interaction effects involving $X_i$ \cite{saltelli_making_2002,saltelli_global_2008}.

The GSA method proposed by Kucherenko \textit{et al.} \cite{kucherenko_estimation_2012} can consider models with dependent input factors. Here, the main and total effects of the variance-based GSA are calculated with a copula-based method. With this approach, $S_i$ includes the effects of the dependence of $X_i$ with other factors \cite{mara_nonparam_2015} and can be higher than $S_{T,i}$. As reported by \cite{mara_nonparam_2015}, $S_{T,i}$ includes only the effects of $X_i$ that are not due to its dependence with $\mathbf{X}_{\sim i}$.
A given factor whose importance is only due to the correlation with another factor would have $S_{T,i}=0$, but $S_i$ can be different from 0 \cite{mara_nonparam_2015}. Moreover, $S_{T,i}$ approaches 0 as the correlation $|\rho| \rightarrow 1$ \cite{kucherenko_estimation_2012}. A possible explanation for this behaviour is that as the correlation approaches 1, the value of $X_i$ is completely informed by $\mathbf{X}_{\sim i}$ and thus $V_{X_i}( Y \, | \, \mathbf{X}_{\sim i} )$ will tend to 0.

\subsection{Latent variable approach for GSA}
This approach expresses the inter-correlation between two parameters as causal relationships between uncorrelated variables. Therefore allowing the use of classical variance-based GSA.

\begin{figure}[t] 
	\centering
	{\includegraphics[scale=0.5]{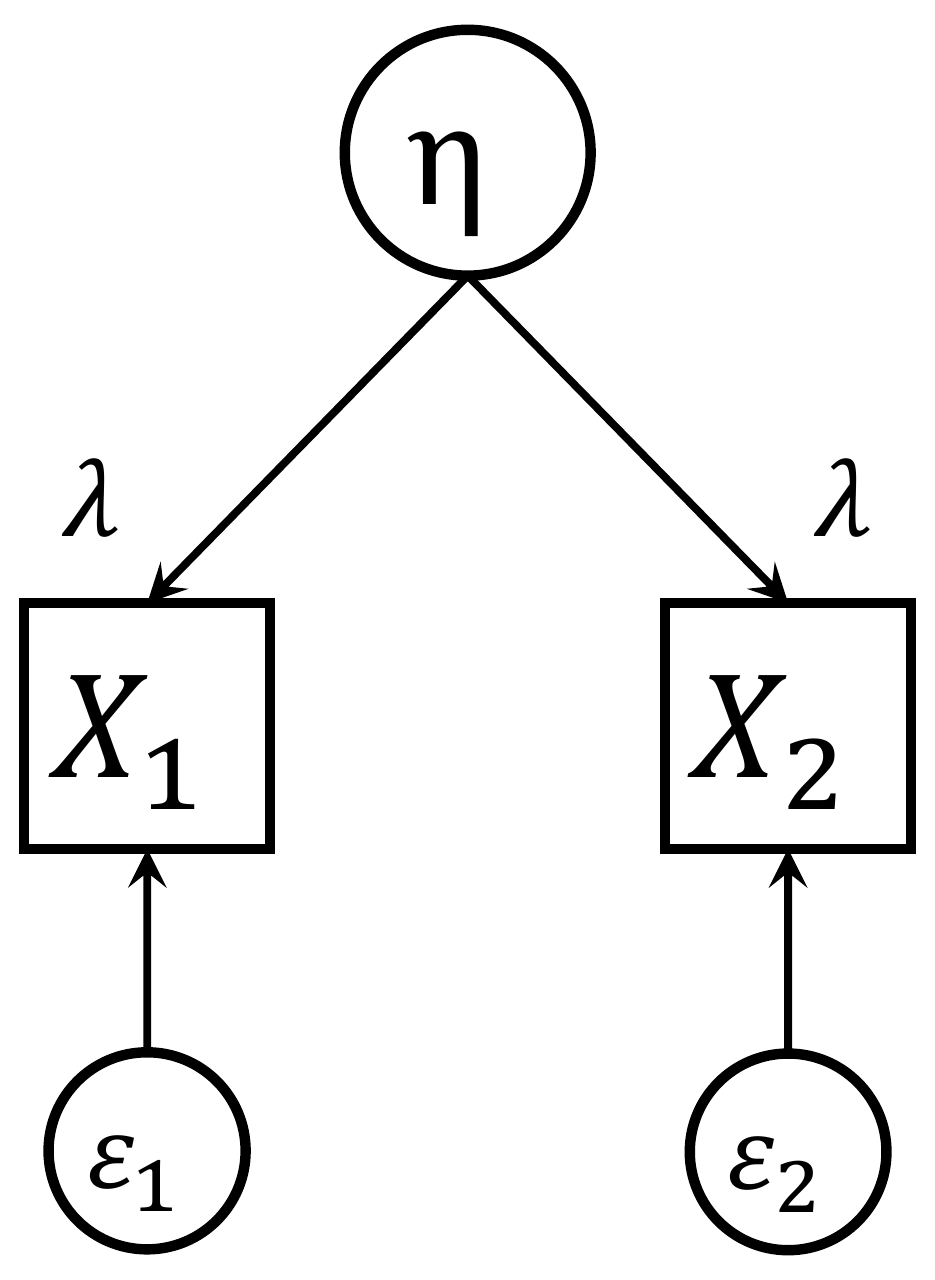}}
	\caption{Relationship between the observed, common and unique variances for two correlated parameters and one latent variable. $X_1$ and $X_2$ are the observed variables, $\eta$ is the latent variable, $\varepsilon_1$ and $\varepsilon_2$ are the unique variances and $\lambda$ are the factor loadings.}
	\label{fig_precursor}
\end{figure}

Latent variable methods partition the \textit{observed variance} of each correlated parameter (observed variable) into two parts: a \textit{common variance}, caused by the latent variable and a \textit{unique variance}, specific to the parameter itself \cite{brown_confirmatory_2015}. In this work, we focus on the case of two linearly correlated random variables whose correlation is explained by one latent variable. The relationship between the observed, common and unique variances for two correlated parameters and one latent variable is reported through a path diagram as shown in Figure \ref{fig_precursor} \cite{loehlin_latent_2017}.
Following the notation of latent-variable methodology, $\eta$ is the latent variable, and is conventionally represented by a circle in the path diagram. Unidirectional arrows represent the causal relationships between latent and dependent factors $X_i$, $i=1,2$ (depicted by a box) and $\varepsilon_i$ represents the unique variance associated with $X_i$ \cite{brown_confirmatory_2015}. $X_1$ and $X_2$ are considered linearly correlated, with a linear (Pearson) correlation coefficient of $\rho_{12}$. Here we assume that $\eta$, $X_i$ and $\varepsilon_i$ are distributed as in Equation system \ref{eq_distr} and that $\eta$ and $\varepsilon_i$ are independent.
\begin{equation}\label{eq_distr}
\begin{aligned}
    \eta &\sim \mathcal{N} (0, 1) \\
    X_i &\sim \mathcal{N} (0,1) \\
    \varepsilon_i &\sim \mathcal{N} (0, \sigma_i^2) 
\end{aligned}
\end{equation}
A common assumption is that the causal relationships between $\eta$ and $X_i$ are linear. In this case, it is possible to write the following Equation system \ref{eq_latent_rel} \cite{loehlin_latent_2017,brown_confirmatory_2015}.
\begin{equation}
\begin{aligned}\label{eq_latent_rel}
    X_1 &= \lambda_1 \, \eta + \varepsilon_1 \\
    X_2 &= \lambda_2 \, \eta + \varepsilon_2
\end{aligned}
\end{equation}
$\lambda_1$ and $\lambda_2$ are called the \textit{factor loadings} and represent the correlations of $X_1$ and $X_2$ with $\eta$ \cite{farrell_ave_2009}. Given that our hypothesis is that $\eta$ and $X_i$ are standard normal random variables, and that $\varepsilon_i$ is distributed normally with a mean equal to 0 and variance $\sigma_i^2$, by calculating the variance of both sides of the equations in Equation system \ref{eq_latent_rel}, it is possible to derive that $\sigma_i^2 = (1 - \lambda_i^2)$, $i=1,2$. 

Now, to correctly express $X_1$ and $X_2$ as functions of $\eta$, we need to define $\lambda_1$, $\lambda_2$ and $\sigma_1^2$, $\sigma_2^2$. According to \textit{path analysis} theory, the correlation between $X_1$ and $X_2$ can be expressed as $\rho_{12} = \lambda_1 \cdot \lambda_2$ \cite{loehlin_latent_2017}. With the hypotheses that $\rho_{12}>0$ and that $X_1$ and $X_2$ have the same relationship with $\eta$, thus $\lambda_1=\lambda_2=\lambda$, it is possible to define $\lambda$ as in Equation \ref{eq_lambda} \cite{loehlin_latent_2017}.
\begin{equation}\label{eq_lambda}
    \lambda = \sqrt{\rho_{12}}
\end{equation}
Another possible solution is $\lambda=-\sqrt{\rho_{12}}$, where the latent variable has a negative correlation with both $X_1$ and $X_2$. In case of $\rho_{12}<0$, the absolute values of both factors loadings are equal to $\sqrt{\rho_{12}}$, while their signs are opposite. 

According to Equation \ref{eq_latent_rel}, $\lambda^2$ is the portion of the variance of $X_i$ that is attributed to the latent factor. With our approach, $\lambda^2$ is the average variance extracted (AVE). AVE can be defined as \textit{`the average amount of variation that a latent construct is able to explain in the observed variables'} \cite{farrell_ave_2009}. Intuitively, this is the overall amount of variance that `is taken' from our dependent factors $X_i$ and attributed to the latent variable $\eta$, in order to define the causal relationships in Equation \ref{eq_latent_rel}.
The general AVE expression, corresponding to one latent variable and $k$ unique variances, is reported in Equation \ref{eq_AVE} \cite{fornell_evaluating_1981}. Considering that  $\sigma_i^2=1-\lambda_i^2$, AVE can be calculated as the average of the squares of the factor loadings associated with the latent variable \cite{farrell_ave_2009}.
\begin{equation}\label{eq_AVE}
    AVE = \frac{ \sum_{i=1}^{k} \lambda_i^2 }{ \sum_{i=1}^{k} \lambda_i^2 + \sum_{i=1}^{k} \sigma_i^2} = \frac{1}{k} \sum_{i=1}^{k} \lambda_i^2
\end{equation}
Considering that in our case $k=2$ (two dependent factors) and $\lambda_1 \cdot \lambda_2 = \rho$, we can derive the expression in Equation \ref{eq_AVE_2}.
\begin{equation}\label{eq_AVE_2}
    AVE = \frac{1}{2}(\lambda_1^2 + \lambda_2^2) = \frac{1}{2}\biggl( \lambda_1^2 + \frac{\rho_{12}^2}{\lambda_1^2} \biggr)
\end{equation}
If we calculate the first derivative of AVE over $\lambda_1$ and set it equal to zero, we can obtain the following expression in Equation \ref{eq_AVE_3}.
\begin{equation}\label{eq_AVE_3}
    \begin{aligned}
        \frac{dAVE}{d\lambda_1} &= \lambda_1 - \frac{\rho_{12}^2}{\lambda_1^3}=0 \\
        \lambda_1 &= \sqrt{|\rho_{12}|} \\
        \lambda_2 &= sign(\rho_{12})\cdot \sqrt{|\rho_{12}|}
    \end{aligned}
\end{equation}
Where, $sign(\rho_{12})$ is equal to +1 if $\rho_{12}>0$, while it is equal to -1 if $\rho_{12}<0$. If we calculate the second derivative we can see that it is always positive, thus $|\lambda_1| = |\lambda_2|$ corresponds to a minimum. With our hypothesis that $X_1$ and $X_2$ have the same relationship with $\eta$, the AVE is minimised. This means that we are explaining the correlation between two observed variables by attributing (on average) the minimum variance possible to the latent construct.

\begin{table}[t]
\captionof{table}{Assumptions for the use of the latent variable approach}
    \centering
    \begin{tabular}{l}
    \toprule
        \textbf{Assumptions$^a$} \\
    \midrule
        Only two correlated input factors $X_1$ and $X_2$\\
        A linear correlation between $X_1$ and $X_2$\\
        $\eta$, $\varepsilon_1$, $\varepsilon_2$, $X_1$, $X_2$ normally distributed as in Equation \ref{eq_distr}\\
        Linear relation between $\eta$ and $X_1$, $X_2$, as in Equation \ref{eq_latent_rel}\\
        Same relation between $X_1$, $X_2$ and $\eta$, thus $|\lambda_1|=|\lambda_2|=|\lambda|$ in Equation \ref{eq_latent_rel}\\
    \bottomrule
    \multicolumn{1}{l}{$^a$ $X_1$, $X_2$ are the dependent input factors} \\
    \multicolumn{1}{l}{ $\eta$ is the latent variable} \\
    \multicolumn{1}{l}{$\varepsilon_1$, $\varepsilon_2$ are the unique variances} \\
    \end{tabular}
    \label{tab:hp_latent}
\end{table}

With the latent variable approach, instead of two correlated random variables ($X_1$ and $X_2$), three independent random variables ($\eta$, $\varepsilon_1$ and $\varepsilon_2$) will be considered in the variance-based GSA. In this context, the impact of $\varepsilon_1$ and $\varepsilon_2$ on the model output can be uniquely attributed to $X_1$ and $X_2$, respectively. Instead, it would be impossible to distinguish if the impact of $\eta$ on the model output is primarily mediated by $X_1$ or $X_2$.

For simplicity, we have considered standardised variables. However, the latent variable approach can easily be extended to data in original units with the use of simple transformations. Nevertheless, in order to use this method several assumptions must be satisfied (summarised in Table \ref{tab:hp_latent}) and some limitations still exist. The sums of the random variables representing the latent and independent variances must follow the distributions of $X_i$. This condition is satisfied if both the parameters are normally distributed and it can easily be extended to the case of the two parameters being log-normally distributed (although in this case $\log(X_1)$ and $\log(X_2)$ must be linearly correlated). However, the condition in Equation system \ref{eq_latent_rel} is not easily satisfied for other types of distributions. The method presented here is valid when considering two correlated factors and it can be extended to three mutually correlated factors, by using the so called \textit{method of triads} to derive a unique solution for the factor loadings \cite{loehlin_latent_2017}. However, it is possible that there is no unique solution when more than three mutually correlated factors are considered \cite{loehlin_latent_2017}. In this situation, the application of the latent variable approach for GSA would become more challenging.

\subsection{Algebraic models}
The latent variable approach was initially tested on three algebraic models, namely model 1, 2 and 3, in Equations \ref{eq_lin_1}, \ref{eq_lin_2} and \ref{eq_lin_3} respectively.
\begin{equation}\label{eq_lin_1}
    Y = X_1 + X_2 + X_2 \cdot X_3
\end{equation}
\begin{equation}\label{eq_lin_2}
    Y = X_1 + X_2 + X_1 \cdot X_3
\end{equation}
\begin{equation}\label{eq_lin_3}
    Y = X_1 + X_2 + X_3 + X_4
\end{equation}
For all the models, all factors were considered to be normally distributed with means equal to 0 and variances equal to 1, $X_i \sim \mathcal{N} (0,1)$, $i=1,2,3,4$. $X_1$ and $X_4$ were considered linearly correlated, with a Pearson correlation coefficient of $\rho_{14}$. Model 1 and model 2 differ in the fact that in model 1, $X_1$ is not involved in any interaction, while in model 2, $X_1$ interacts with $X_3$.\\
$X_4$ does not appear in the model 1 or model 2 equations, consequently, its `causal impact'\footnote{Here we refer to `causal impact' as the impact of an input factor $X_i$ on the model output $Y$ that is not due to the dependence of $X_i$ with other factors.} on the model output $Y$ must be null. Intuitively, for both model 1 and 2, the results of a variance-based GSA in absence of correlation, considering only $X_1$, $X_2$ and $X_3$, will correctly reflect the structure of the model.

\subsection{Whole-body PBPK model for midazolam}
\begin{figure}[h] 
	\centering
	{\includegraphics[scale=0.9]{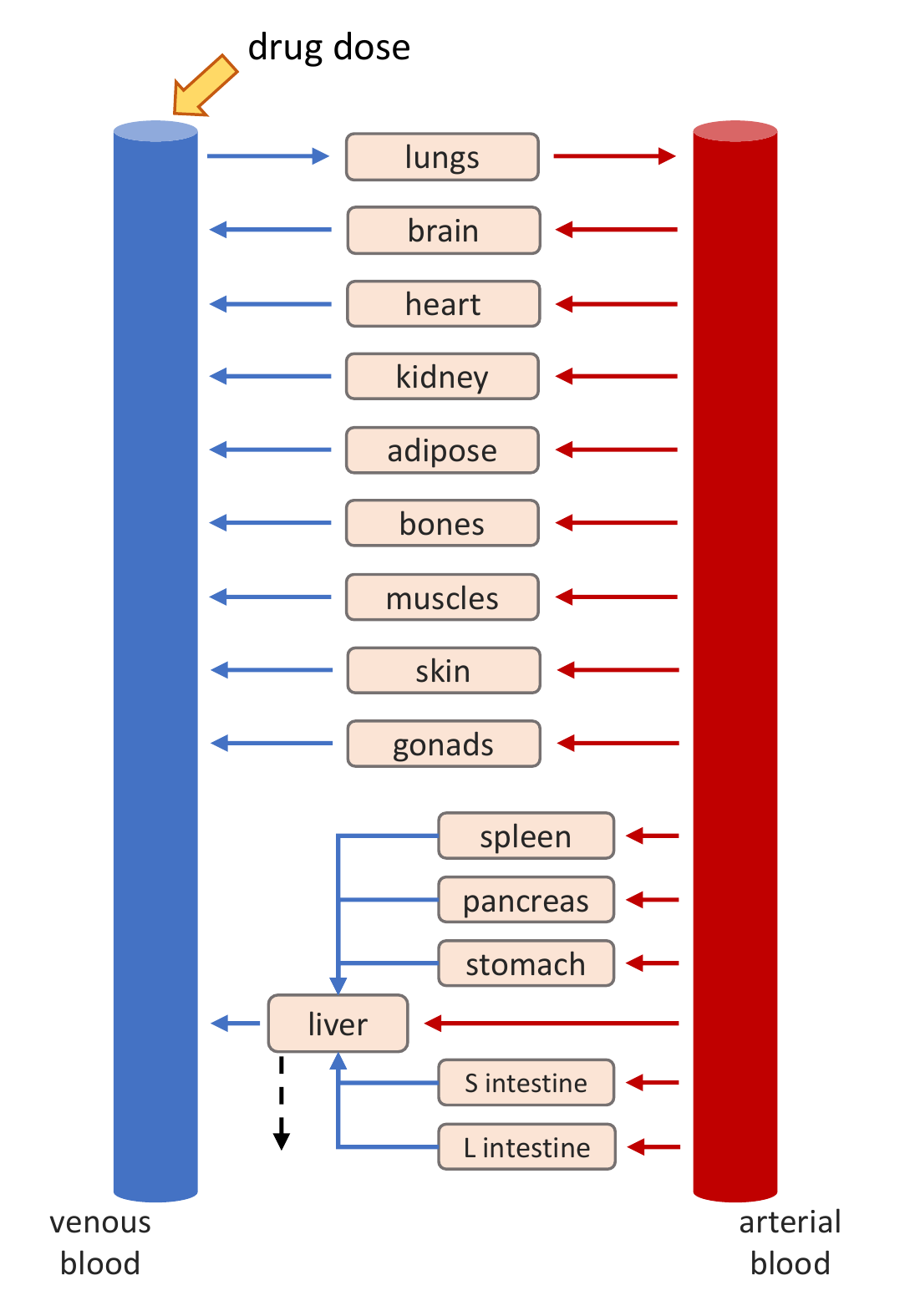}}
	\caption{Structure of a general whole-body PBPK model. Each box corresponds to a specific compartment. The red and blue arrows represent the arterial and venous blood flow, respectively. The black-dashed arrow represents elimination through metabolism in the liver. The yellow arrow represent the drug intravenous administration. \textit{S intestine} and \textit{L intestine} are the small and large intestine, respectively.}
	\label{fig_pbpk_struct}
\end{figure}

A whole-body PBPK model was developed, describing the pharmacokinetics of the drug MDZ following an intravenous (IV) bolus injection in a population of human healthy subjects. The model is represented in Figure \ref{fig_pbpk_struct}. This section provides a brief description of the model. For a detailed account of the model equations, the parameters used for the PBPK construction and the algorithm used for generating the population, see the Supplementary Material.

The typical equation used to describe the mass balance in a given organ or tissue $t$ within a PBPK model is reported in Equation \ref{eq_pbpk_t}. For a detailed description and the underlying theories of this model, called \textit{well-stirred perfusion-limited} PBPK, please refer to \cite{berezhkovskiy_valid_2010}.
\begin{equation}\label{eq_pbpk_t}
    \frac{dx_t}{dt} = Q_t \, \biggl( \frac{x_{art}}{V_{art}} - \frac{x_t/V_t}{P_{t:p}/B:P} \biggr)
\end{equation}
Equation \ref{eq_pbpk_t} is valid for all organs and tissues except the liver, the lungs, the arterial and venous blood. $x_t$ is the drug amount in compartment $t$, while $V_t$ is the volume. Subscript $art$ stands for arterial blood. $Q_t$ is the blood flow to compartment $t$. $B:P$ is the blood-to-plasma ratio and $P_{t:p}$ is the tissue-to-plasma partition coefficient.

MDZ is primarily metabolised in the liver by the two enzymes, CYP3A4 and CYP3A5. For MDZ both enzymes catalyse two reactions, leading to the formation of two metabolites,\textit{1-hydroxy} \textit{midazolam} (1-OH-MDZ) and \textit{4-hydroxy} \textit{midazolam} (4-OH-MDZ) \cite{vossen_dynamically_2007,Galetin_midazolam_2004}. For this reason, two mass flows corresponding to MDZ metabolism leave the PBPK system from the liver compartment, as represented in Equation system \ref{eq_liver}.
\begin{equation}\label{eq_liver}
    \begin{aligned}
    \frac{dx_{liv}}{dt} &= Q_{liv} \biggl(\frac{x_{art}}{V_{art}} - \frac{x_{liv}/V_{liv}}{P_{liv:p}/B:P}\biggr) + \sum_{t \in \mathcal{S}} \Biggl[ Q_t \, \biggl( \frac{x_t/V_t}{P_{t:p}/B:P} \biggr) \Biggr] \\
    & \, - MET_{3A4} - MET_{3A5}
    \end{aligned}
\end{equation}
Subscript $liv$ stands for liver, $\mathcal{S}$ represents the splanchnic organs (spleen, pancreas, stomach, small and large intestine). $c_{u,liv}$ is the unbound liver concentration. $MET_{3A4}$ and $MET_{3A5}$ are the fluxes representing the reactions catalysed by CYP3A4 and CYP3A5.  All the chemical reactions are described using \textit{Michaelis-Menten} equations \cite{michaelis_kinetik_1913}. The \textit{Michaelis-Menten} parameters for MDZ are taken from \textit{in vitro} studies \cite{Galetin_midazolam_2004} and they are scaled to the \textit{in vivo} context as per \cite{rostamihodjegan_simulation_2007}. One of the main parameters used for the \textit{in vitro} to \textit{in vivo} scaling is the microsomal protein per gram of liver ($MPPGL$) (see supplementary materials for a detailed description of this process).

The population variability of physiological parameters such as the compartment volumes and blood flow was generated with a simple algorithm having as inputs the sex of the subject, the height and the body mass index (BMI).

To simulate an IV bolus injection of 5 $mg$ of MDZ, the initial condition of the venous blood compartment was set equal to 5, while the remaining compartments were set to equal 0. The \textit{area under the curve} (AUC) of the venous plasma compartment was considered the output of interest for the GSA. The AUC is defined as in Equation \ref{eq_AUC}.
\begin{equation}\label{eq_AUC}
    AUC = \int_{t_{in}}^{t_{end}} \frac{x_{ven}(\tau)}{V_{ven} \cdot B:P} \,d\tau
\end{equation}
$t_{in}$ and $t_{end}$ were set to 0 and $24\cdot 7$ $h$, respectively. The AUC is a measure of the cumulative exposure of a drug over time. In PK, AUC is an important metric not only to represent exposure, but also to calculate a number of PK parameters using non-compartmental analysis \cite{rowland_clinical_1995}. The distributions of the model parameters considered in this analysis are reported in Table \ref{tab_pbpk_param}.

\begin{table}[h]
	\captionof{table}{Variable parameters used for the MDZ PBPK model}
	\centering
	\resizebox{\textwidth}{!}{
	\begin{tabular}[ht]{ l c c c c}
		\toprule
		\textbf{Parameters} & distribution parameters & distribution type & units & references \\
		\midrule
		sex$^{d}$               & 0, 1            & uniform$^a$     &           &                               \\
		height (male)$^{e}$     & 176.7 (6.15)    & normal$^b$      & $cm$      & \cite{cacciari_italian_2006}  \\
		height (female)$^{e}$   & 163.3 (5.85)    & normal$^b$      & $cm$      & \cite{cacciari_italian_2006}  \\
		BMI$^{f}$               & 18.5, 24.9      & uniform$^a$     & $kg/m^2$  & \cite{WHO:BMI}                  \\
		$[CYP3A4]^g$              & 137 (41\%)      & log-normal$^c$  & $(pmol\,CYP)/(mg\,MP)$  & \cite{cubitt_sources_2011}\\
		$[CYP3A5]^g$              & 103 (65\%)      & log-normal$^c$  & $(pmol\,CYP)/(mg\,MP)$  & \cite{cubitt_sources_2011}\\
		$MPPGL^h$                 & 39.79 (27\%)    & log-normal$^c$  & $(mg\,prot)/(g\,liver)$ &  \cite{noauthor_simcyp_2017}\\
		\bottomrule
		\multicolumn{5}{l}{$^{a}$ For distribution parameters, \textit{minimum}, \textit{maximum} of the parameter.}  \\
		\multicolumn{5}{l}{$^{c}$ For distribution parameters, \textit{mean (standard deviation)} of the normal variable.} \\
		\multicolumn{5}{l}{$^{c}$ For distribution parameters, \textit{mean (CV)} of the log-normal variable.} \\
		\multicolumn{5}{l}{$^{d}$ If the extracted value is $<0.5$ the subject is female (0), otherwise male (1).} \\
		\multicolumn{5}{l}{$^{e}$ height for a 20 years old Italian population} \\
		\multicolumn{5}{l}{$^{f}$ Body mass index corresponding to the nutritional status of `Normal weight' according to the World Health Organization} \\
		\multicolumn{5}{l}{$^{g}$ CYP abundance per mg of microsomal protein} \\
		\multicolumn{5}{l}{$^{h}$ mg of microsomal proteins for gram of liver} \\
		\multicolumn{5}{l}{} \\
	\end{tabular}}
	\label{tab_pbpk_param}
\end{table}

\subsection{Analysis overview}
For the GSA, the following methods were applied to both the algebraic and the PBPK models:
\begin{itemize}
    \item classical variance-based GSA considering all the parameters uncorrelated;
    \item variance-based GSA grouping together the two correlated parameters;
    \item the method developed by Kucherenko for computing the variance-based GSA indices in presence of correlation \cite{kucherenko_estimation_2012};
    \item the latent variable approach.
\end{itemize}

Concerning the algebraic models, the analysis was carried out varying $\rho_{14}$, from -0.9 to 0.9. When $\rho_{14}>0$, the latent variable was considered to be positively correlated with both $X_1$ and $X_4$ ($\lambda>0$). Instead, when $\rho_{14}<0$, the latent variable was considered to be positively correlated with $X_1$ and negatively correlated with $X_4$.

For the PBPK model, the (Pearson) correlation between the logarithms of CYP3A4 and CYP3A5 abundances $\rho_{3A4,3A5}$ was considered to equal 0.52, based on proteomic data from human liver samples \cite{achour_simultaneous_2014}, for the variance-based GSA with grouped factors, for the Kucherenko and the latent variable approaches. In this analysis, all simulated individuals were assumed to express CYP3A5.

All analysis was performed in MATLAB R2020a\footnote{The codes are made available at the following link \url{https://github.com/NicolaMelillo/latent_variable_GSA}.} \cite{noauthor_matlab_2019}. The systems of differential equations were solved with the \texttt{ode15s} MATLAB solver, for a timespan ranging from 0 to $24\cdot 7$ $h$. GSA was performed using the software UQLab \cite{Marelli_uqlab_2014} except for the variance-based GSA with groups, where an \textit{`ad hoc'} MATLAB code was developed. For the numerical estimation of the sensitivity indices, within UQLab, the \textit{`homma'} estimator was used for the Sobol approach, while the default estimator embedded in the software was used for the Kucherenko approach. Concerning the \textit{`ad hoc'} MATLAB code, we used the estimator reported in \cite{saltelli_global_2008} (the \textit{errata corrige} version).
For all the methods, the sample size was fixed to 10,000. The uncertainty of the sensitivity indices estimates was assessed by using 1,000 bootstrap samples, with the exception of the Kucherenko method, where the convergence plots were used.
\clearpage
\section{Results}
\label{section_results}
\subsection{Algebraic models}
\begin{figure}[h] 
	\centering
	{\includegraphics[scale=0.4]{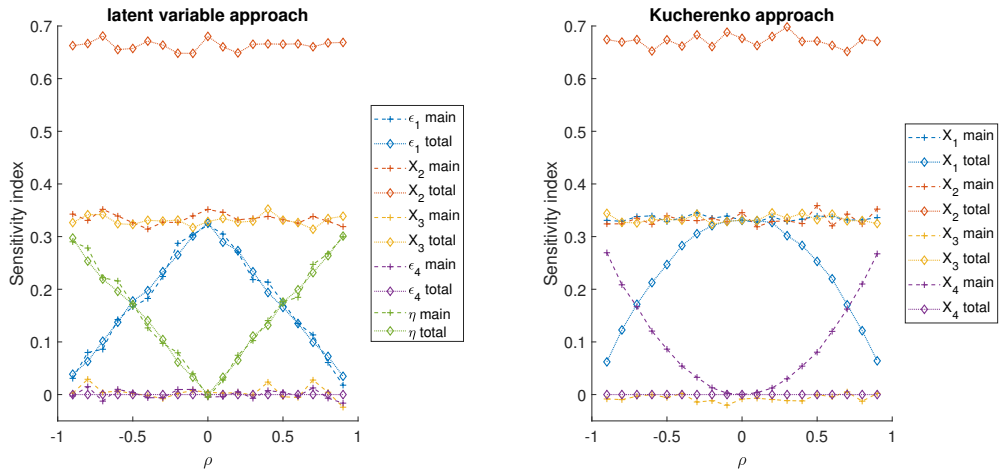}}
	\caption{Algebraic model 1 GSA results of the latent variable and the method presented by Kucherenko 2012 \cite{kucherenko_estimation_2012}.}
	\label{fig_res_model1}
\end{figure}

\begin{figure}[h] 
	\centering
	{\includegraphics[scale=0.4]{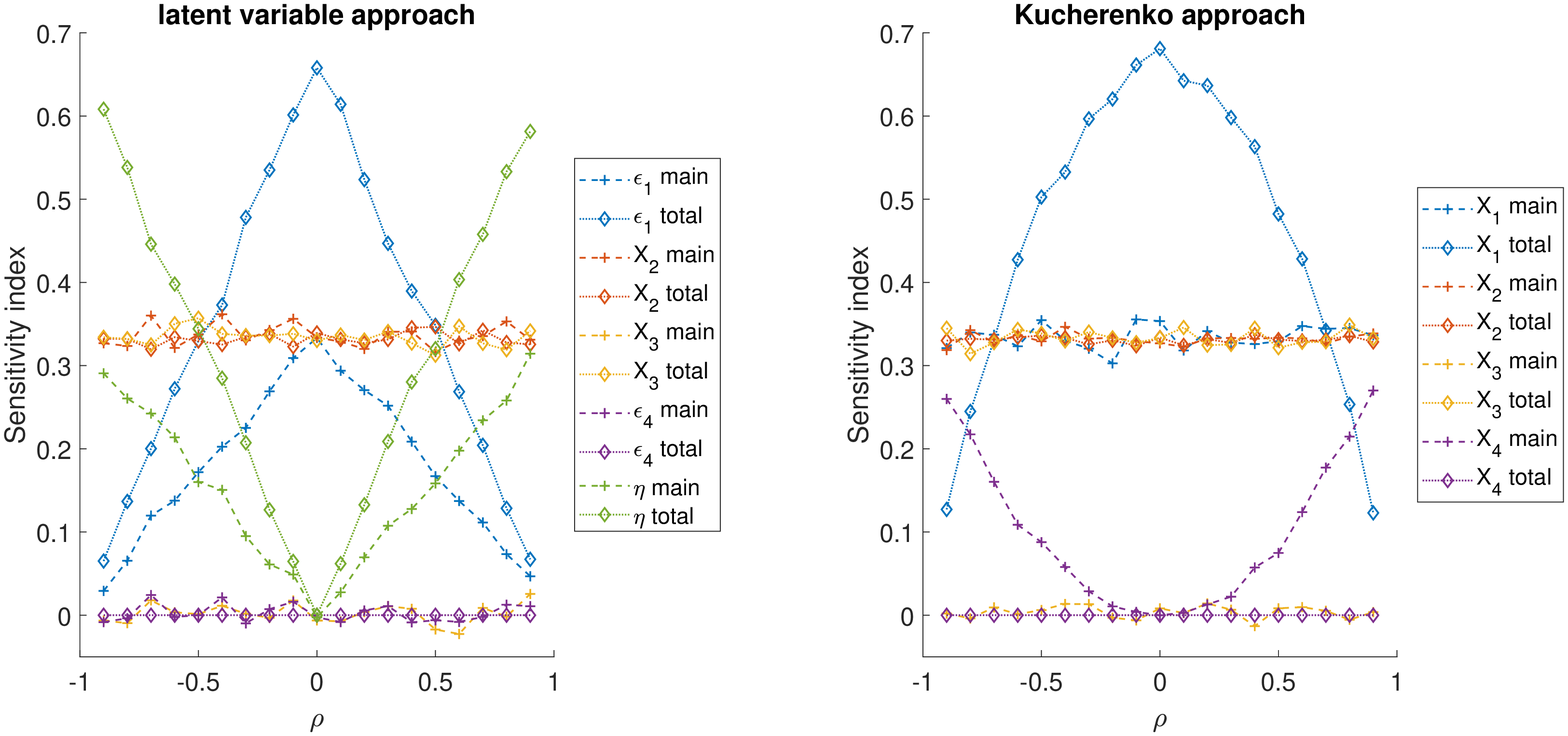}}
	\caption{Algebraic model 2 GSA results of the latent variable and the method presented by Kucherenko 2012 \cite{kucherenko_estimation_2012}.}
	\label{fig_res_model2}
\end{figure}

\begin{figure}[h] 
	\centering
	{\includegraphics[scale=0.4]{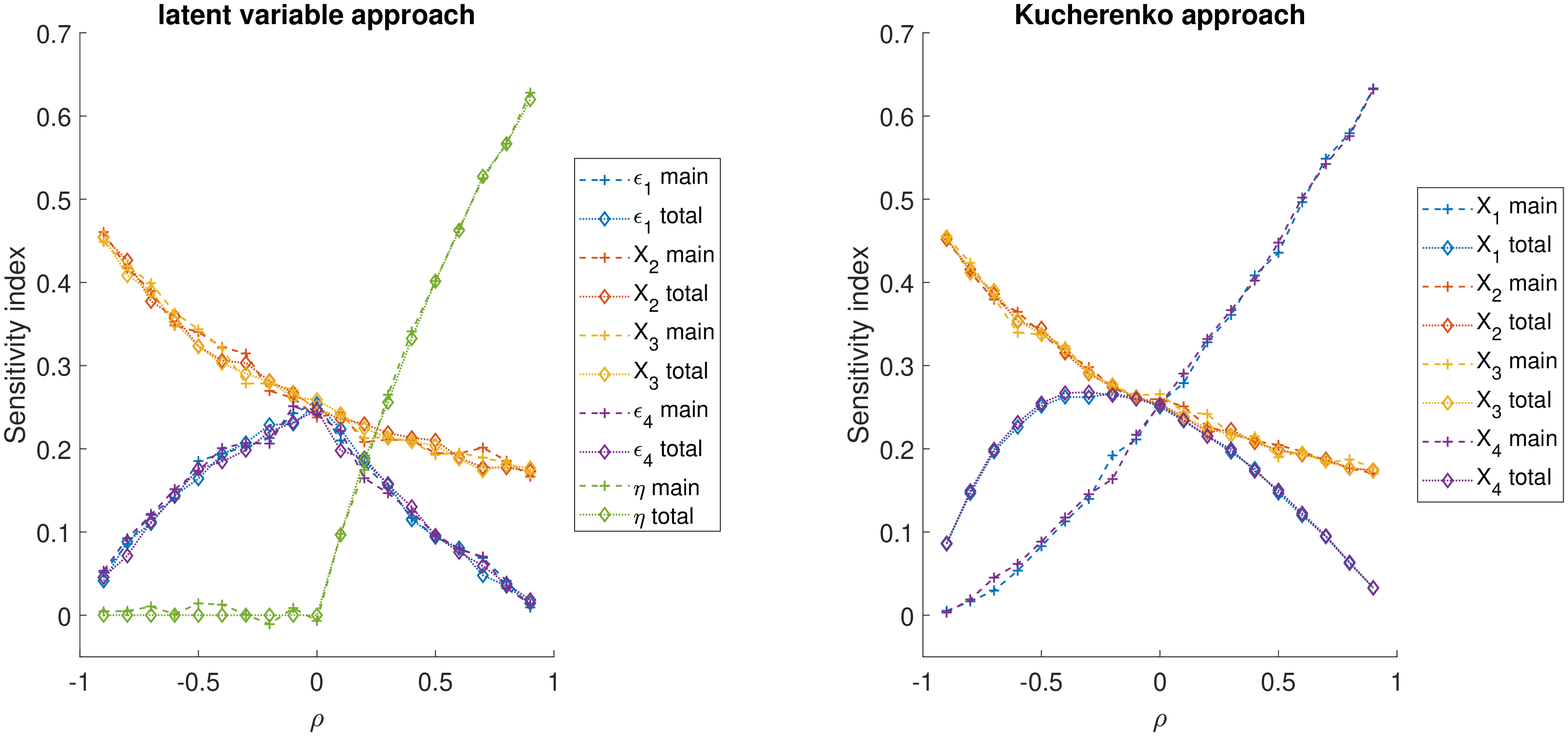}}
	\caption{Algebraic model 3 GSA results of the latent variable and the method presented by Kucherenko 2012 \cite{kucherenko_estimation_2012}.}
	\label{fig_res_model3}
\end{figure}

The GSA results for the algebraic models 1, 2 and 3, with $\rho_{14}=0.7$ and $\rho_{14}=0.9$, are reported in Tables \ref{table_model_1}, \ref{table_model_2} and \ref{table_model_3}, respectively. In Figure \ref{fig_res_model1} the GSA results obtained with the latent variable and the Kucherenko approaches for the algebraic model 1 are given as a function of $\rho_{14}$, ranging from -0.9 to 0.9. For the models 2 and 3, the equivalent information is shown in Figure \ref{fig_res_model2} and \ref{fig_res_model3}, respectively. Here we begin by reporting the results of model 1 and 2 and then, model 3.

The parameter $X_4$ does not appear in Equations \ref{eq_lin_1} and \ref{eq_lin_2}. Regardless of presence or absence of correlation between $X_1$ and $X_4$ its `causal' impact on the output should therefore be null. Hence, intuitively, the results of a variance-based GSA with the classic Sobol's method considering only $X_1$, $X_2$ and $X_3$ should be the ones that truly represent the model structure. Any differences in main and total effects for the Kucherenko approach, the latent variable approach and the variance based GSA with grouped factors are therefore due to how these methods handle the correlation.

Concerning the Kucherenko approach, in Figure \ref{fig_res_model1} the higher the absolute value of $\rho_{14}$ is, the higher the main effect of $X_4$ is, while its total effect always remains equal to 0. This substantially confirms the findings of \cite{mara_nonparam_2015}. Moreover, as the absolute value of the correlation increases, the total effect of $X_1$ decreases, while the main effect remains stable. From \cite{mara_nonparam_2015} we know that $S_1$ includes the impact of the correlation of $X_1$ with $X_4$, while $S_{T,1}$ just includes the `uncorrelated' effects. From our example is possible to appreciate that the higher $|\rho_{14}|$ is, the lower the `uncorrelated' effect of $X_1$ is. In this context it is actually challenging to distinguish between the `causal' effect of $X_1$ and $X_4$ on $Y$ and the effect due to their dependence. Similar conclusions can be made for the model 2. By limiting the analysis to the Kucherenko indices, it is challenging to understand how much $X_1$ is involved in interaction effects and, ultimately, to determine any ranking of importance of the parameters as can be used in practical applications.

Concerning the latent variable approach, presented in Figures \ref{fig_res_model1} and \ref{fig_res_model2}, the higher the absolute value of $\rho_{14}$ is, the higher the importance of the latent variable over the unique variances. Ultimately, with $\rho_{14}$ approaching 1 the whole variance of both $X_1$ and $X_4$ becomes fully explained by the latent factor and thus, the residual variances' effect on the output variance tends to 0. Given that the latent variable affects both the correlated factors equally, it is not possible to elucidate if the impact of $\eta$ on the output variance is primarily mediated by $X_1$ or $X_4$. However, the impact of the unique variances can be uniquely attributed to the correlated factors. In fact, for both models 1 and 2, both the main and total effect of $\varepsilon_4$ are always equal to zero, as seen in Figures \ref{fig_res_model1} and \ref{fig_res_model2}. This is unlikely the case for traditional variance-based GSA with groups (see Tables \ref{table_model_1} and \ref{table_model_2}), where, independently of the values of $\rho_{14}$, it is not possible to determine the impact of the variable within the groups. Notably, if $|\rho|$ is close to 1, the latent variable will fully explain both $X_1$ and $X_4$, resembling the case of the grouping approach.
Given that in both the grouping and the latent variable approach we are performing a standard Sobol's GSA with uncorrelated factors, the interpretation of the sensitivity indices and the factor ranking is straightforward.

In model 1, $X_1$ is not involved in any interactions. This is discernible when $S_i=S_{T,i}$. In this case, $S_1=S_{T,1}$, as seen in Table \ref{table_model_1} and Figure \ref{fig_res_model2}. Neither $\eta$ or $\varepsilon_1$ are involved in any interactions. This is quite intuitive as the model is linear and $X_1$ is defined as the sum of the latent variable and the unique variance in the latent variable approach. However, interaction effects between the latent variable and the unique variance will arise, for example, in case of $X_1$ having a nonlinear effect (\textit{e.g.}, quadratic) on $Y$\footnote{If $X_1=\lambda \eta + \varepsilon$ and $Y=X_1^2$, it is straightforward to derive that $Y=\lambda^2 \, \eta^2 + \varepsilon^2 + 2 \lambda \eta \varepsilon$. In this case, there are interaction effects between $\eta$ and $\varepsilon$.}. In model 2, $X_1$ and $X_3$ show interaction effects, as noted in the Sobol's GSA results. This happens when $S_{T,i}>S_i$. In Table \ref{table_model_2} and Figure \ref{fig_res_model2} we can see that both the latent variable and the unique variance of $X_1$ show interaction effects.

Concerning model 3, Table \ref{table_model_3} and Figure \ref{fig_res_model3}, we observe that the sensitivity indices of $X_2$ and $X_3$ change in function of $\rho_{14}$. The traditional variance-based GSA that considers all the factors uncorrelated does not capture this effect. With this simple example, we can see that ignoring the correlation within GSA could potentially bias the overall results of the analysis. Traditional GSA with groups can capture this effect and thus, it can be an easy an reliable method for treating correlations. However, as explained for models 1 and 2, it has the limitation of not distinguishing the impact of the variables within the groups of correlated factors.

Concerning the Kucherenko approach, $S_1$ and $S_4$ are close to 0 when $\rho_{14}$ is close to 0 and they both grow as $|\rho_{14}|$ grows. Instead, $S_{T,1}$ and $S_{T,2}$ have almost a parabolic shape. Both the main and total effects of $X_1$ and $X_4$ are low for strong negative correlation, probably because in this model the effect of $X_1$ tends to cancel the one of $X_4$ on $Y$ and vice versa. For a high positive correlation the total effects tend to zero, while the main effects are close to 0.6. 

Regarding the latent variable approach, one interesting observation is that the overall tendency of the unique variances and latent variable sensitivity indices are similar to those of the total and main effects of $X_1$ and $X_4$ of the Kucherenko approach, respectively. This probably happens because the unique variances represents the impact of the `uncorrelated' part of the factors, similarly to the total effect of the Kucherenko approach. Instead, both the latent variable and the main effect include the `dependent' part of the factors. However, one important difference is that the latent variable approach is a variance-based GSA performed with independent variables and thus, the indices are easily understandable, this is unlikely the case for the Kucherenko approach. Finally, it is interesting to observe that for negative correlations the impact of the latent variable is zero. This happens because the factor loadings ($\lambda$) are equal in module, but opposite in sign and thus, the latent variable term is cancelled from Equation \ref{eq_lin_3}.

\begin{table}
    \captionof{table}{Results for the algebraic model 1}
	\centering
	\resizebox{\textwidth}{!}{
    \begin{tabular}[h]{lcccccccc}
        \toprule
        & \multicolumn{2}{c}{{Sobol$^a$}} & \multicolumn{2}{c}{{Kucherenko$^b$}} & \multicolumn{2}{c}{{Latent variable$^a$}} & \multicolumn{2}{c}{{Grouped$^a$}} \\
        \cmidrule{2-9}
        Factor & main & total & main & total & main & total & main & total \\
        \midrule
        &  & & \multicolumn{6}{c}{{$\rho_{14}=0.7$}} \\
        \cmidrule{4-9}
        $X_1^c$                & \begin{tabular}[c]{@{}c@{}}0.34\\ (0.32,0.36)\end{tabular} & \begin{tabular}[c]{@{}c@{}}0.33\\ (0.31,0.34)\end{tabular} & 0.33 & 0.17 & \begin{tabular}[c]{@{}c@{}}0.11\\ (0.09,0.13)\end{tabular}  & \begin{tabular}[c]{@{}c@{}}0.1\\ (0.9,0.11)\end{tabular}   & \begin{tabular}[c]{@{}c@{}}0.31$^d$\\ (0.29,0.33)\end{tabular} & \begin{tabular}[c]{@{}c@{}}0.34$^d$\\ (0.31,0.37)\end{tabular} \\
        $X_2$                & \begin{tabular}[c]{@{}c@{}}0.33\\ (0.31,0.35)\end{tabular} & \begin{tabular}[c]{@{}c@{}}0.67\\ (0.64,0.7)\end{tabular}  & 0.32 & 0.64 & \begin{tabular}[c]{@{}c@{}}0.32\\ (0.3,0.35)\end{tabular}   & \begin{tabular}[c]{@{}c@{}}0.65\\ (0.63,0.67)\end{tabular} & \begin{tabular}[c]{@{}c@{}}0.31\\ (0.28,0.33)\end{tabular} & \begin{tabular}[c]{@{}c@{}}0.7\\ (0.67,0.72)\end{tabular}  \\
        $X_3$                & \begin{tabular}[c]{@{}c@{}}0\\ (-0.03,0.02)\end{tabular}   & \begin{tabular}[c]{@{}c@{}}0.33\\ (0.31,0.35)\end{tabular} & 0    & 0.34 & \begin{tabular}[c]{@{}c@{}}0.02\\ (-0.01,0.04)\end{tabular} & \begin{tabular}[c]{@{}c@{}}0.33\\ (0.31,0.35)\end{tabular} & \begin{tabular}[c]{@{}c@{}}-0.03\\ (-0.06,0)\end{tabular}  & \begin{tabular}[c]{@{}c@{}}0.32\\ (0.29,0.34)\end{tabular} \\
        $X_4^c$                & \begin{tabular}[c]{@{}c@{}}0\\ (-0.02,0.02)\end{tabular}   & \begin{tabular}[c]{@{}c@{}}0\\ (0,0)\end{tabular}          & 0.16 & 0    & \begin{tabular}[c]{@{}c@{}}0.02\\ (0,0.03)\end{tabular}     & \begin{tabular}[c]{@{}c@{}}0\\ (0,0)\end{tabular}          &                                                            &                                                            \\
        $\eta$ &   &    &      &      & \begin{tabular}[c]{@{}c@{}}0.26\\ (0.24,0.28)\end{tabular}  & \begin{tabular}[c]{@{}c@{}}0.23\\ (0.22,0.25)\end{tabular} &        & \\                                             
        & & & \multicolumn{6}{c}{{$\rho_{14}=0.9$}} \\
        \cmidrule{4-9}
        
        $X_1^c$ & \begin{tabular}[c]{@{}c@{}}0.33\\ (0.31,0.35)\end{tabular}   & \begin{tabular}[c]{@{}c@{}}0.35\\ (0.33,0.37)\end{tabular} & 0.33          & 0.06         & \begin{tabular}[c]{@{}c@{}}0.05\\ (0.03,0.07)\end{tabular}  & \begin{tabular}[c]{@{}c@{}}0.04\\ (0.03,0.04)\end{tabular} & \begin{tabular}[c]{@{}c@{}}0.33$^d$\\ (0.31,0.35)\end{tabular} & \begin{tabular}[c]{@{}c@{}}0.34$^d$\\ (0.31,0.37)\end{tabular} \\
        $X_2$ & \begin{tabular}[c]{@{}c@{}}0.32\\ (0.29,0.34)\end{tabular}   & \begin{tabular}[c]{@{}c@{}}0.66\\ (0.64,0.69)\end{tabular} & 0.33          & 0.69         & \begin{tabular}[c]{@{}c@{}}0.33\\ (0.31,0.35)\end{tabular}  & \begin{tabular}[c]{@{}c@{}}0.65\\ (0.63,0.68)\end{tabular} & \begin{tabular}[c]{@{}c@{}}0.35\\ (0.33,0.38)\end{tabular} & \begin{tabular}[c]{@{}c@{}}0.67\\ (0.64,0.7)\end{tabular}  \\
        $X_3$ & \begin{tabular}[c]{@{}c@{}}-0.01\\ (-0.04,0.02)\end{tabular} & \begin{tabular}[c]{@{}c@{}}0.33\\ (0.31,0.36)\end{tabular} & -0.01         & 0.35         & \begin{tabular}[c]{@{}c@{}}0.02\\ (-0.01,0.04)\end{tabular} & \begin{tabular}[c]{@{}c@{}}0.35\\ (0.33,0.37)\end{tabular} & \begin{tabular}[c]{@{}c@{}}0\\ (-0.03,0.02)\end{tabular}   & \begin{tabular}[c]{@{}c@{}}0.33\\ (0.31,0.36)\end{tabular} \\
        $X_4^c$ & \begin{tabular}[c]{@{}c@{}}-0.01\\ (-0.03,0.01)\end{tabular} & \begin{tabular}[c]{@{}c@{}}0\\ (0,0)\end{tabular}          & 0.27          & 0            & \begin{tabular}[c]{@{}c@{}}0.01\\ (-0.01,0.03)\end{tabular} & \begin{tabular}[c]{@{}c@{}}0\\ (0,0)\end{tabular}          &                                                            &                                                            \\
        $\eta$  &   &     &  &   & \begin{tabular}[c]{@{}c@{}}0.3\\ (0.28,0.32)\end{tabular}   & \begin{tabular}[c]{@{}c@{}}0.29\\ (0.27,0.3)\end{tabular}  &    &      \\                        
        \bottomrule 
        \multicolumn{9}{l}{$^a$ values reported in the table are mean (2.5,97.5) percentiles calculated with 1000 bootstrap samples} \\
        \multicolumn{9}{l}{$^b$ convergence plots are shown in the supplementary materials} \\
        \multicolumn{9}{l}{$^c$ for the latent variable model, this refers to the unique variance} \\
         \multicolumn{9}{l}{$^d$ this refers to the $X_1$ and $X_4$ group} \\
    \end{tabular}}
    \label{table_model_1}
\end{table}

\begin{table}
    \captionof{table}{Results for the algebraic model 2}
	\centering
	\resizebox{\textwidth}{!}{
    \begin{tabular}[h]{lcccccccc}
        \toprule
        & \multicolumn{2}{c}{{Sobol$^a$}} & \multicolumn{2}{c}{{Kucherenko$^b$}} & \multicolumn{2}{c}{{Latent variable$^a$}} & \multicolumn{2}{c}{{Grouped$^a$}} \\
        \cmidrule{2-9}
        Factor & main & total & main & total & main & total & main & total \\
        \midrule
        &  & & \multicolumn{6}{c}{{$\rho_{14}=0.7$}} \\
        \cmidrule{4-9}
        
        $X_1^c$ & \begin{tabular}[c]{@{}c@{}}0.34\\ (0.32,0.36)\end{tabular}   & \begin{tabular}[c]{@{}c@{}}0.68\\ (0.66,0.71)\end{tabular} & 0.32 & 0.34 & \begin{tabular}[c]{@{}c@{}}0.11\\ (0.09,0.13)\end{tabular}  & \begin{tabular}[c]{@{}c@{}}0.2\\ (0.18,0.21)\end{tabular}  & \begin{tabular}[c]{@{}c@{}}0.33$^d$\\ (0.3,0.35)\end{tabular}     & \begin{tabular}[c]{@{}c@{}}0.68$^d$\\ (0.65,0.71)\end{tabular} \\
        $X_2$ & \begin{tabular}[c]{@{}c@{}}0.32\\ (0.3,0.34)\end{tabular}    & \begin{tabular}[c]{@{}c@{}}0.33\\ (0.32,0.35)\end{tabular} & 0.33 & 0.32 & \begin{tabular}[c]{@{}c@{}}0.33\\ (0.31,0.35)\end{tabular}  & \begin{tabular}[c]{@{}c@{}}0.33\\ (0.31,0.35)\end{tabular} & \begin{tabular}[c]{@{}c@{}}0.32\\ (0.3,0.34)\end{tabular}     & \begin{tabular}[c]{@{}c@{}}0.34\\ (0.31,0.37)\end{tabular} \\
        $X_3$ & \begin{tabular}[c]{@{}c@{}}0\\ (-0.02,0.03)\end{tabular}     & \begin{tabular}[c]{@{}c@{}}0.34\\ (0.32,0.36)\end{tabular} & 0    & 0.34 & \begin{tabular}[c]{@{}c@{}}0.01\\ (-0.01,0.04)\end{tabular} & \begin{tabular}[c]{@{}c@{}}0.33\\ (0.3,0.35)\end{tabular}  & \begin{tabular}[c]{@{}c@{}}-0.03\\ (-0.05,-0.01)\end{tabular} & \begin{tabular}[c]{@{}c@{}}0.33\\ (0.31,0.35)\end{tabular} \\
        $X_4^c$ & \begin{tabular}[c]{@{}c@{}}-0.01\\ (-0.03,0.01)\end{tabular} & \begin{tabular}[c]{@{}c@{}}0\\ (0,0)\end{tabular}          & 0.16 & 0    & \begin{tabular}[c]{@{}c@{}}0.01\\ (-0.01,0.02)\end{tabular} & \begin{tabular}[c]{@{}c@{}}0\\ (0,0)\end{tabular}          &                                                               &                                                            \\
        $\eta$  &  & & & & \begin{tabular}[c]{@{}c@{}}0.24\\ (0.22,0.26)\end{tabular}  & \begin{tabular}[c]{@{}c@{}}0.47\\ (0.45,0.49)\end{tabular} &  &  \\

        & & & \multicolumn{6}{c}{{$\rho_{14}=0.9$}} \\
        \cmidrule{4-9}
        
        $X_1^c$ & \begin{tabular}[c]{@{}c@{}}0.33\\ (0.31,0.35)\end{tabular} & \begin{tabular}[c]{@{}c@{}}0.66\\ (0.63,0.69)\end{tabular} & 0.32 & 0.13 & \begin{tabular}[c]{@{}c@{}}0.03\\ (0.01,0.05)\end{tabular}   & \begin{tabular}[c]{@{}c@{}}0.06\\ (0.05,0.07)\end{tabular} & \begin{tabular}[c]{@{}c@{}}0.36$^d$\\ (0.33,0.38)\end{tabular}  & \begin{tabular}[c]{@{}c@{}}0.65$^d$\\ (0.62,0.68)\end{tabular} \\
        $X_2$ & \begin{tabular}[c]{@{}c@{}}0.32\\ (0.3,0.34)\end{tabular}  & \begin{tabular}[c]{@{}c@{}}0.34\\ (0.32,0.35)\end{tabular} & 0.32 & 0.33 & \begin{tabular}[c]{@{}c@{}}0.32\\ (0.3,0.34)\end{tabular}    & \begin{tabular}[c]{@{}c@{}}0.33\\ (0.32,0.35)\end{tabular} & \begin{tabular}[c]{@{}c@{}}0.35\\ (0.33,0.37)\end{tabular}  & \begin{tabular}[c]{@{}c@{}}0.32\\ (0.29,0.35)\end{tabular} \\
        $X_3$ & \begin{tabular}[c]{@{}c@{}}0\\ (-0.03,0.03)\end{tabular}   & \begin{tabular}[c]{@{}c@{}}0.34\\ (0.32,0.37)\end{tabular} & 0    & 0.35 & \begin{tabular}[c]{@{}c@{}}-0.01\\ (-0.03,0.01)\end{tabular} & \begin{tabular}[c]{@{}c@{}}0.34\\ (0.32,0.37)\end{tabular} & \begin{tabular}[c]{@{}c@{}}0.01\\ (-0.01,0.04)\end{tabular} & \begin{tabular}[c]{@{}c@{}}0.34\\ (0.32,0.37)\end{tabular} \\
        $X_4^c$ & \begin{tabular}[c]{@{}c@{}}0\\ (-0.02,0.02)\end{tabular}   & \begin{tabular}[c]{@{}c@{}}0\\ (0,0)\end{tabular}          & 0.25 & 0    & \begin{tabular}[c]{@{}c@{}}0\\ (-0.02,0.02)\end{tabular}     & \begin{tabular}[c]{@{}c@{}}0\\ (0,0)\end{tabular}          &                                                             &                                                            \\
        $\eta$  &                                                            &                                                            &      &      & \begin{tabular}[c]{@{}c@{}}0.29\\ (0.27,0.32)\end{tabular}   & \begin{tabular}[c]{@{}c@{}}0.61\\ (0.59,0.64)\end{tabular} &                                                             &                                                            \\
        \bottomrule 
        \multicolumn{9}{l}{$^a$ values reported in the table are mean (2.5,97.5) percentiles calculated with 1000 bootstrap samples} \\
        \multicolumn{9}{l}{$^b$ convergence plots are shown in the supplementary materials} \\
        \multicolumn{9}{l}{$^c$ for the latent variable model, this refers to the unique variance} \\
         \multicolumn{9}{l}{$^d$ this refers to the $X_1$ and $X_4$ group} \\
    \end{tabular}}
    \label{table_model_2}
\end{table}

\begin{table}
    \captionof{table}{Results for the algebraic model 3}
	\centering
	\resizebox{\textwidth}{!}{
    \begin{tabular}[h]{lcccccccc}
        \toprule
        & \multicolumn{2}{c}{{Sobol$^a$}} & \multicolumn{2}{c}{{Kucherenko$^b$}} & \multicolumn{2}{c}{{Latent variable$^a$}} & \multicolumn{2}{c}{{Grouped$^a$}} \\
        \cmidrule{2-9}
        Factor & main & total & main & total & main & total & main & total \\
        \midrule
        &  & & \multicolumn{6}{c}{{$\rho_{14}=0.7$}} \\
        \cmidrule{4-9}
        $X_1^c$                & \begin{tabular}[c]{@{}c@{}}0.25\\ (0.23,0.26)\end{tabular} & \begin{tabular}[c]{@{}c@{}}0.25\\ (0.23,0.26)\end{tabular} & 0.55 & 0.1  & \begin{tabular}[c]{@{}c@{}}0.07\\ (0.05,0.09)\end{tabular} & \begin{tabular}[c]{@{}c@{}}0.05\\ (0.04,0.06)\end{tabular} & \begin{tabular}[c]{@{}c@{}}0.62$^d$\\ (0.6,0.64)\end{tabular}  & \begin{tabular}[c]{@{}c@{}}0.63$^d$\\ (0.61,0.65)\end{tabular} \\
        $X_2$                & \begin{tabular}[c]{@{}c@{}}0.25\\ (0.23,0.26)\end{tabular} & \begin{tabular}[c]{@{}c@{}}0.24\\ (0.23,0.26)\end{tabular} & 0.19 & 0.19 & \begin{tabular}[c]{@{}c@{}}0.18\\ (0.16,0.2)\end{tabular}  & \begin{tabular}[c]{@{}c@{}}0.19\\ (0.18,0.2)\end{tabular}  & \begin{tabular}[c]{@{}c@{}}0.19\\ (0.17,0.21)\end{tabular} & \begin{tabular}[c]{@{}c@{}}0.18\\ (0.16,0.2)\end{tabular}  \\
        $X_3$                & \begin{tabular}[c]{@{}c@{}}0.26\\ (0.24,0.27)\end{tabular} & \begin{tabular}[c]{@{}c@{}}0.25\\ (0.24,0.27)\end{tabular} & 0.18 & 0.19 & \begin{tabular}[c]{@{}c@{}}0.2\\ (0.18,0.22)\end{tabular}  & \begin{tabular}[c]{@{}c@{}}0.19\\ (0.18,0.2)\end{tabular}  & \begin{tabular}[c]{@{}c@{}}0.18\\ (0.16,0.2)\end{tabular}  & \begin{tabular}[c]{@{}c@{}}0.19\\ (0.17,0.2)\end{tabular}  \\
        $X_4^c$                & \begin{tabular}[c]{@{}c@{}}0.26\\ (0.24,0.28)\end{tabular} & \begin{tabular}[c]{@{}c@{}}0.25\\ (0.24,0.27)\end{tabular} & 0.54 & 0.1  & \begin{tabular}[c]{@{}c@{}}0.06\\ (0.04,0.08)\end{tabular} & \begin{tabular}[c]{@{}c@{}}0.05\\ (0.05,0.06)\end{tabular} &                                                            &                                                            \\
        $\eta$ &                                                            &                                                            &      &      & \begin{tabular}[c]{@{}c@{}}0.51\\ (0.49,0.53)\end{tabular} & \begin{tabular}[c]{@{}c@{}}0.51\\ (0.49,0.53)\end{tabular} &                                                            &                                                            \\

        & & & \multicolumn{6}{c}{{$\rho_{14}=0.9$}} \\
        \cmidrule{4-9}
        
        $X_1^c$                & \begin{tabular}[c]{@{}c@{}}0.24\\ (0.22,0.26)\end{tabular} & \begin{tabular}[c]{@{}c@{}}0.25\\ (0.24,0.26)\end{tabular} & 0.63 & 0.03 & \begin{tabular}[c]{@{}c@{}}0.02\\ (0,0.04)\end{tabular}    & \begin{tabular}[c]{@{}c@{}}0.02\\ (0.02,0.02)\end{tabular} & \begin{tabular}[c]{@{}c@{}}0.65$^d$\\ (0.63,0.67)\end{tabular} & \begin{tabular}[c]{@{}c@{}}0.65$^d$\\ (0.63,0.67)\end{tabular}  \\
        $X_2$                & \begin{tabular}[c]{@{}c@{}}0.24\\ (0.22,0.26)\end{tabular} & \begin{tabular}[c]{@{}c@{}}0.25\\ (0.24,0.26)\end{tabular} & 0.17 & 0.17 & \begin{tabular}[c]{@{}c@{}}0.18\\ (0.16,0.2)\end{tabular}  & \begin{tabular}[c]{@{}c@{}}0.17\\ (0.16,0.18)\end{tabular} & \begin{tabular}[c]{@{}c@{}}0.17\\ (0.15,0.19)\end{tabular} & \begin{tabular}[c]{@{}c@{}}0.17\\ (0.15,0.19)\end{tabular}  \\
        $X_3$                & \begin{tabular}[c]{@{}c@{}}0.26\\ (0.24,0.28)\end{tabular} & \begin{tabular}[c]{@{}c@{}}0.24\\ (0.23,0.25)\end{tabular} & 0.18 & 0.17 & \begin{tabular}[c]{@{}c@{}}0.17\\ (0.15,0.19)\end{tabular} & \begin{tabular}[c]{@{}c@{}}0.17\\ (0.16,0.18)\end{tabular} & \begin{tabular}[c]{@{}c@{}}0.18\\ (0.16,0.2)\end{tabular}  & \begin{tabular}[c]{@{}c@{}}0.19\\  (0.17,0.21)\end{tabular} \\
        $X_4^c$                & \begin{tabular}[c]{@{}c@{}}0.25\\ (0.23,0.27)\end{tabular} & \begin{tabular}[c]{@{}c@{}}0.26\\ (0.25,0.28)\end{tabular} & 0.62 & 0.03 & \begin{tabular}[c]{@{}c@{}}0.02\\ (0,0.04)\end{tabular}    & \begin{tabular}[c]{@{}c@{}}0.02\\ (0.01,0.02)\end{tabular} &                                                            &                                                             \\
        $\eta$ &                                                            &                                                            &      &      & \begin{tabular}[c]{@{}c@{}}0.63\\ (0.61,0.64)\end{tabular} & \begin{tabular}[c]{@{}c@{}}0.62\\ (0.6,0.64)\end{tabular}  &                                                            &                                                             \\

        \bottomrule 
        \multicolumn{9}{l}{$^a$ values reported in the table are mean (2.5,97.5) percentiles calculated with 1000 bootstrap samples} \\
        \multicolumn{9}{l}{$^b$ convergence plots are shown in the supplementary materials} \\
        \multicolumn{9}{l}{$^c$ for the latent variable model, this refers to the unique variance} \\
         \multicolumn{9}{l}{$^d$ this refers to the $X_1$ and $X_4$ group} \\
    \end{tabular}}
    \label{table_model_3}
\end{table}

\subsection{Whole-body PBPK model for midazolam}
\begin{figure}[h] 
	\centering
	{\includegraphics[scale=0.5]{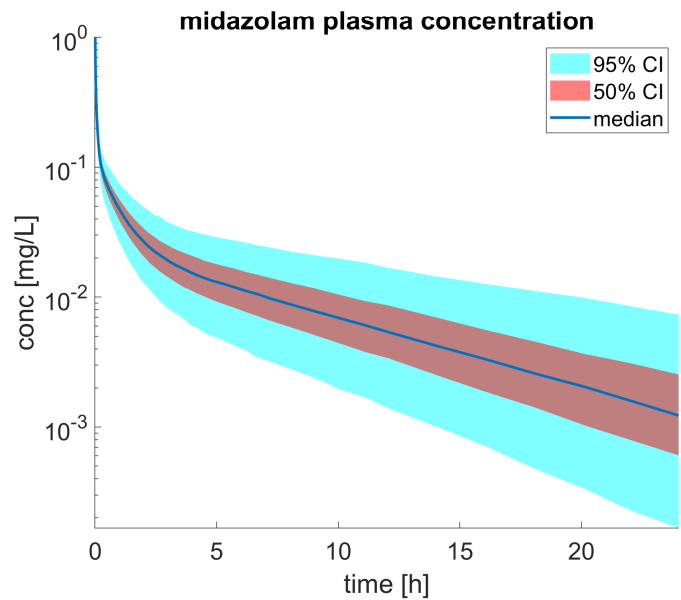}}
	\caption{Simulated population midazolam plasma concentration over time following an intravenous (IV) bolus dose of 5 mg. The simulation was performed with the PBPK model for 10,000 individuals. The physiological correlation was considered between the abundances of CYP3A4 and CYP3A5.}
	\label{fig_plasma_conc}
\end{figure}

\begin{figure}[h] 
	\centering
	{\includegraphics[scale=0.5]{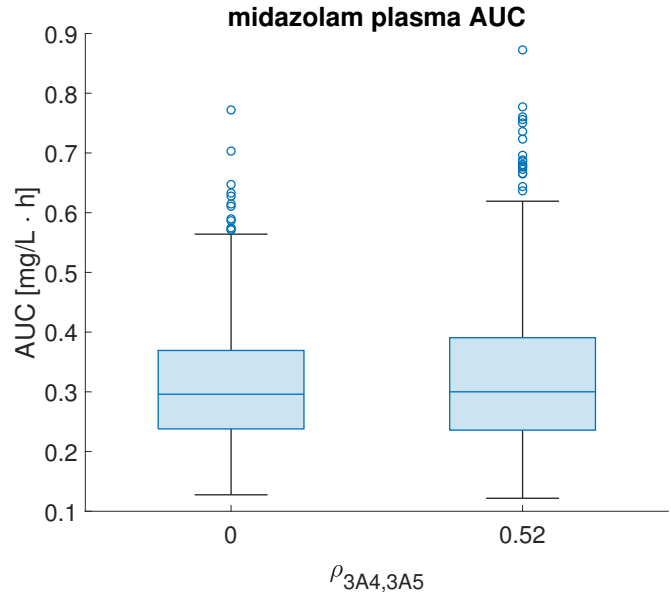}}
	\caption{Simulated midazolam AUC distribution both in presence and absence of correlation between CYP3A4 and CYP3A5 abundances. The simulation was performed with the PBPK model for 10,000 individuals.}
	\label{fig_boxplot}
\end{figure}

The simulated MDZ plasma concentration-time profiles and AUCs for a population of 10,000 subjects are shown in Figures \ref{fig_plasma_conc} and \ref{fig_boxplot}, respectively. The GSA results of Sobol's method without accounting for the correlation, of the Kucherenko method, of the traditional variance-based GSA with groups and of the latent variable approach are presented in Table \ref{table_pbpk_res}. 

According to the results from Sobol's GSA, the most important parameters in explaining the variability in AUC are (in order of importance) the MPPGL, CYP3A4 and CYP3A5 abundances. These factors are important because they control the rate of metabolism in the liver. The fact that the metabolism-related parameters are the most important for explaining variability in AUC suggests that the rate-limiting step of drug elimination is the metabolism and not, for example, liver blood flow. Given that exposure drives drug effect, the interindividual variability in efficacy, due to PK, is mainly explained by genetics in this case example. However, we need to consider that our population is composed by healthy adults with a BMI corresponding to the nutritional status of `normal weight' \cite{WHO:BMI}. The inclusion of overweight or obese subjects may impact the results of the GSA.

Concerning the GSA results obtained with the Kucherenko, the variance-based GSA with groups and the latent variable approach, the sensitivity indices of MPPGL are slightly reduced as compared to Sobol's GSA. This is most likely related with the fact that the correlation between CYP3A4 and CYP3A5 tends to generate more `extreme' individuals, \textit{i.e.}, poor metabolisers (with low CYP3A4 and low CYP3A5 abundances) and rapid metabolisers (with high CYP3A4 and high CYP3A5 abundances). Thus, as it is possible to observe in Figure \ref{fig_boxplot}, the AUC distribution in case of correlation is slightly wider with respect to the case of no correlation. These results are in agreement with our previous studies \cite{melillo_accounting_2019}.

Concerning the Kucherenko analysis, it is difficult to confidently use either the main or the total effects for the purpose of factor ranking. For example, by observing the main effect the two most important parameters are CYP3A4 and CYP3A5 abundances. However, it is difficult to understand what the contributions of the variables themselves are and what is due to the correlation. For this reason, in our example, there is a risk of overestimating the importance of the enzymatic abundances and, by extension, underestimating the importance of the other factors. By using the total effect for the factor ranking, there is instead the risk of underestimating the importance of the correlated factors and overestimating the importance of the remaining inputs, as the total effects for the factors involved in the correlation tend to 0 as $|\rho| \rightarrow 1$ \cite{kucherenko_estimation_2012}. Moreover, by using these two indices, given that for both CYP3A4 and CYP3A5 abundances the total effect is lower than the main effect, it is difficult to understand the effect of interactions.

In the latent variable approach, the factor ranking can be done by examining either the main or at the total effects. This is possible because the correlation between CYP3A4 and CYP3A5 was expressed in terms of a functional relationship between three independent factors, the latent variable and two independent variances. Thus, the classical variance-based GSA was used. With this approach, the most important factor in explaining the AUC is $\eta$, followed by MPPGL and the independent components of CYP3A4 and CYP3A5. By using either the main or the total effect for the factor ranking, we can confidently assess that the main drivers for the plasma AUC are the metabolism-related parameters. Moreover, with this method it is possible to appreciate the interaction effects, that in this case are mild and do not have a great impact on the factor ranking. A downside of this approach is that $\eta$ drives both CYP3A4 and CYP3A5 variability. For this reason, given that the latent variable is one of the two most important parameters, it is not possible to appreciate if its importance is primarily caused by the CYP3A4 or CYP3A5 mediated pathway. By investigating the independent components of CYP3A4 and CYP3A5 abundances, it is noted that they do have a similar impact. Intuitively, if one of the two factors was not important for the AUC, the independent component would be equal to zero (however, it is not necessarily true for the opposite case).

The results of the PBPK simulations presented here aim to illustrate a GSA methodology, only. Therefore, we do not recommend their use for other purposes.

\begin{table} 
    \captionof{table}{GSA results for the MDZ PBPK model}
	\centering
	\resizebox{\textwidth}{!}{
        \begin{tabular}{lcccccccc}
        \toprule
        & \multicolumn{2}{c}{{Sobol$^a$}} & \multicolumn{2}{c}{{Kucherenko$^b$}} & \multicolumn{2}{c}{{Latent variable$^a$}} & \multicolumn{2}{c}{{Grouped$^a$}}\\
        \cmidrule{2-9}
        Factor & main & total & main & total & main & total & main & total \\
        \midrule
         & & & \multicolumn{6}{c}{{$\rho_{3A4,3A5}=0.52$}} \\
        \cmidrule{4-9}
            sex    & \begin{tabular}[c]{@{}c@{}}0\\ (-0.02,0.02)\end{tabular}    & \begin{tabular}[c]{@{}c@{}}0.02\\ (0.01,0.03)\end{tabular} & 0.01 & 0.02 & \begin{tabular}[c]{@{}c@{}}0.03\\ (0.01,0.05)\end{tabular} & \begin{tabular}[c]{@{}c@{}}0.02\\ (0.01,0.02)\end{tabular} & \begin{tabular}[c]{@{}c@{}}0\\ (-0.02,0.02)\end{tabular}                   & \begin{tabular}[c]{@{}c@{}}0.01\\ (-0.03,0.04)\end{tabular}                    \\
            height & \begin{tabular}[c]{@{}c@{}}0.01\\ (-0.01,0.03)\end{tabular} & \begin{tabular}[c]{@{}c@{}}0.05\\ (0.04,0.05)\end{tabular} & 0.02 & 0.03 & \begin{tabular}[c]{@{}c@{}}0.04\\ (0.02,0.06)\end{tabular} & \begin{tabular}[c]{@{}c@{}}0.03\\ (0.02,0.04)\end{tabular} & \begin{tabular}[c]{@{}c@{}}0.01\\ (-0.01,0.03)\end{tabular}                & \begin{tabular}[c]{@{}c@{}}0.01\\ (-0.02,0.05)\end{tabular}                 \\
            BMI    & \begin{tabular}[c]{@{}c@{}}0.03\\ (0.01,0.05)\end{tabular}  & \begin{tabular}[c]{@{}c@{}}0.05\\ (0.04,0.06)\end{tabular} & 0.03 & 0.03 & \begin{tabular}[c]{@{}c@{}}0.04\\ (0.02,0.07)\end{tabular} & \begin{tabular}[c]{@{}c@{}}0.03\\ (0.02,0.05)\end{tabular} & \begin{tabular}[c]{@{}c@{}}0.01\\ (-0.01,0.03)\end{tabular}                & \begin{tabular}[c]{@{}c@{}}0.03\\ (-0.01,0.06)\end{tabular}                 \\
            MPPGL  & \begin{tabular}[c]{@{}c@{}}0.29\\ (0.27,0.31)\end{tabular}  & \begin{tabular}[c]{@{}c@{}}0.39\\ (0.37,0.41)\end{tabular} & 0.25 & 0.3  & \begin{tabular}[c]{@{}c@{}}0.26\\ (0.24,0.29)\end{tabular} & \begin{tabular}[c]{@{}c@{}}0.3\\ (0.27,0.32)\end{tabular}  & \begin{tabular}[c]{@{}c@{}}0.24\\ (0.22,0.27)\end{tabular}                  & \begin{tabular}[c]{@{}c@{}}0.29\\ (0.26,0.32)\end{tabular}                 \\
            CYP3A4$^c$ & \begin{tabular}[c]{@{}c@{}}0.27\\ (0.25,0.3)\end{tabular}   & \begin{tabular}[c]{@{}c@{}}0.33\\ (0.31,0.35)\end{tabular} & 0.49 & 0.22 & \begin{tabular}[c]{@{}c@{}}0.12\\ (0.1,0.15)\end{tabular}  & \begin{tabular}[c]{@{}c@{}}0.15\\ (0.13,0.17)\end{tabular} & \multirow{2}{*}{\begin{tabular}[c]{c@{}c@{}}0.61$^b$\\ (0.58,0.64)\end{tabular}} & \multirow{2}{*}{\begin{tabular}[c]{c@{}c@{}}0.69$^d$\\ (0.67,0.72)\end{tabular}} \\
            CYP3A5$^c$ & \begin{tabular}[c]{@{}c@{}}0.23\\ (0.2,0.25)\end{tabular}   & \begin{tabular}[c]{@{}c@{}}0.29\\ (0.27,0.31)\end{tabular} & 0.42 & 0.15 & \begin{tabular}[c]{@{}c@{}}0.09\\ (0.07,0.09)\end{tabular} & \begin{tabular}[c]{@{}c@{}}0.1\\ (0.09,0.12)\end{tabular}  &  &   \\
            $\eta$    &                                                           &                                                         &     &     & \begin{tabular}[c]{@{}c@{}}0.43\\ (0.41,0.46)\end{tabular} & \begin{tabular}[c]{@{}c@{}}0.48\\ (0.46-0.5)\end{tabular}  &                                                                            &      \\                                                                     
        \bottomrule 
        \multicolumn{9}{l}{$^a$ values reported in the table are mean (2.5,97.5) percentiles calculated with 1000 bootstrap samples} \\
        \multicolumn{9}{l}{$^b$ convergence plots are shown in the supplementary materials} \\
        \multicolumn{9}{l}{$^c$ for the latent variable model, this refers to the unique variance} \\
        \multicolumn{9}{l}{$^d$ refers to the group of CYP3A4 and CYP3A5} \\
        \end{tabular}}
\label{table_pbpk_res}
\end{table}
\clearpage
\section{Discussion}\label{section_discussion}

GSA is gaining use in modelling for pharmaceutics, especially in the field of PBPK M\&S. Recent applications in the literature \cite{melillo_variance_2019,melillo_accounting_2019,melillo_inter-compound_2020,mcnally_workflow_2011,zhang_sobol_2015,daga_physiologically_2018-1,yau_gsappkrr_2020,liu_considerations_2020} and regulatory discussions \cite{chmp_ema_guideline_2018,fda_pbpkguide_2018} have indicated the usefulness of these methods and it is likely that GSA will become an important feature of modelling in pharmaceutical R\&D and for regulatory decision-making. This development is welcomed, indeed in the field of toxicology GSA is an important part of best practices for risk assessment of dose metric predictions \cite{mcnally_workflow_2011,meek_case_2013,IPCS_2010}. 

In order for GSA to gain wider use, the issues of usability and interpretation of the results need to be considered. PBPK M\&S is an interdisciplinary effort highly reliant on experts in several domains, including medicinal chemistry, \textit{in vitro} drug metabolism, pharmacokinetics, pharmacology, toxicology, statistical and mathematical modelling, and more. Further, modelling activities are an important tool for supporting a wide variety of decisions in R\&D and regulatory submissions. For this reason, dedicated user-friendly software platforms are widely used, facilitating standardisation and easy access for non-expert users. We suspect that this is likely to hold true across many different domains, and therefore relevant across areas of application. In this context, particular attention in communicating GSA results should be paid.

Most whole-body PBPK models include several sets of correlated parameters, many of which constrain the models to realistic parameter combinations. It is therefore important that these correlations are accounted for when performing GSA. Several GSA methodologies have been proposed to account for dependent inputs \cite{kucherenko_estimation_2012,xu_extending_2007,tarantola_variance-based_2017,mara_nonparam_2015,do_correlation_2020} and the method developed by Kucherenko was applied to PBPK models and implemented in a recent version of one of the most widely used PBPK software platforms in pharmaceutical industry \cite{liu_investigating_2019,noauthor_simcyp_2019}. However, considerable debate is still ongoing amongst GSA practitioners on how to appropriately interpret the outcomes of these methods. We believe that the use of methodologies whose interpretation is still a matter of debate, require appropriate care in cases where GSA is called upon to support critical decisions, such as those relating to patient safety.

In this work, we propose a relatively simple method using a latent variable approach that deals with correlated input variables in variance-based GSA. The method expresses the correlation between two factors as causal relationships between a latent factor, $\eta$, and two unique variances. As a result this allows the use of classical Sobol's GSA with uncorrelated factors. In our opinion, the approach provides an intuitive process for implementation and interpretation as illustrated in the analysis for MDZ. By ranking the factors according to the total effects of Sobol's GSA, it was possible to clearly interpret the sensitivity indices. This allows insights into the model behaviour and to understand what the main drivers of variability are in a given output. By having a unique, easy and universally recognised interpretation of the sensitivity indices, it is possible to use GSA for supporting decision-making with increased confidence.

One of several alternatives to the latent variable approach would be the use of traditional variance-based GSA with groups. The main advantage is that this method allows treating more than two, or three, dependent factors and other dependencies than the linear correlations. However, as highlighted in the results section, with this approach is not possible to separately distinguish the impact of the dependent variables within a given group. Another alternative could be to assign a unidirectional dependency between the two correlated factors as we have done in a previous study in the context of PBPK models \cite{melillo_accounting_2019}. However, by doing so to describe the dependency, this will affect the relative significance of one input over the other. The potentially arbitrary choice of assigning dependency will increase the importance of the independent variable in the GSA and may produce misleading results. With the latent variable approach we renounce any attempt to completely distinguish the impact of the two correlated inputs on a given model output. Instead, we highlight the impact of the latent variable $\eta$ (as the `common cause') along with the independent part.

Here we also attempt to examine the shortcomings of the latent variable approach. In fact, the method presents some limitations with regards to the number and the distribution of the factors that are mutually correlated, as described in section \ref{section_methods}. Moreover, the results of the latent variable approach need to be interpreted in light of the assumptions summarised in Table \ref{tab:hp_latent}. In case one or more of these assumption are not satisfied, the use of traditional GSA with groups is likely a better choice. Despite this, we believe that the latent variable approach can be of use. This would be true at least until further research is done and a clear, universally recognised interpretation of the sensitivity indices have been agreed for more general GSA methods for dependent inputs that rely on fewer assumptions, such as the approaches proposed by Kucherenko \textit{et al.} \cite{kucherenko_estimation_2012} and Mara \textit{et al.} \cite{mara_nonparam_2015}.

\section{Declaration of Interest}
None.

\section{Acknowledgements}
We would like to thank Professor Paolo Magni of Universit\`a degli Studi di Pavia for valuable discussions and suggestions. This research did not receive any specific grant from funding agencies in the public, commercial, or not-for-profit sectors.


\clearpage
\appendix
\section{Physiologically based pharmacokinetic (PBPK) model for midazolam}

The typical equation used to describe the mass balance in a given organ or tissue $t$ within a physiologically based pharmacokinetic (PBPK) model is reported in Equation \ref{eq_pbpk_t}.
\begin{equation}\label{eq_pbpk_t}
    \frac{dx_t}{dt} = Q_t \, \biggl( \frac{x_{art}}{V_{art}} - \frac{x_t/V_t}{P_{t:p}/B:P} \biggr)
\end{equation}
Equation \ref{eq_pbpk_t} is valid for all organs or tissues except the liver, the lungs, the arterial and venous blood. $x_t$ is the drug amount in compartment $t$, while $V_t$ is the volume. Subscript $art$ denotes arterial blood. $Q_t$ is the blood flow to compartment $t$. $B:P$ is the blood-to-plasma ratio, that is an experimentally derived parameter representing the whole blood drug concentration, divided by the plasma drug concentration at steady state. $P_{t:p}$ is the tissue-to-plasma partition coefficient and represents the tissue drug concentration divided by the plasma drug concentration at steady state. Given the challenges in the experimental measurements of this parameter, several semi-empirical models have been developed over the years, describing $P_{t:p}$ as a function of drug and tissue properties \cite{graham_partition_2012}. In this work, the Berezhkovskiy model, given in Equation \ref{eq_berezh}, was used \cite{berezhkovskiy_volume_2004}:

\begin{equation}\label{eq_berezh}
P_{t:p} = \frac{  D_{v,ow}\cdot(V_{nl,t} + 0.3\cdot V_{ph,t}) + (V_{w,t}/fu_t + 0.7\cdot V_{ph,t}) }{ D_{v,ow}\cdot(V_{nl,p} + 0.3\cdot V_{ph,p}) + (V_{w,p}/fu_p + 0.7\cdot V_{ph,p}) }
\end{equation}
$V_{nl,t}$ and $V_{nl,p}$ are the volume fractions of neutral lipids in tissue and plasma, respectively; $V_{ph,t}$ and $V_{ph,p}$ are the volume fractions of phospholipids in tissue and plasma; $V_{w,t}$ and $V_{w,p}$ are the water volume fractions in tissue and plasma. Volume fractions are reported in Table \ref{tab_organs_comp}. $D_{v,ow}$ is the drug partition coefficient between vegetable oil and water and it was obtained as follows $\log{D_{v,ow}}=1.115\cdot \log{P_{o,w}}$, with $\log{P_{o,w}}$ the octanol to water partition coefficient \cite{poulin_prediction_2002}. $fu_p$ and $fu_t$ are the drug fraction unbound in plasma and tissue, with the latter calculated as: $fu_t = 1/( 1 + 0.5 \cdot (1-fu_p)/fu_p )$ \cite{poulin_prediction_2002}. All the drug related parameters are given in Table \ref{tab_drug_param}.

The equations for the lungs, arterial and venous blood are reported in equation system \ref{eq_lung_blood}.
\begin{equation}\label{eq_lung_blood}
    \begin{aligned}
    \frac{dx_{lungs}}{dt} &= Q_{tot} \, \biggl( \frac{x_{ven}}{V_{ven}} - \frac{x_{lungs}/V_{lungs}}{P_{lungs:p}/B:P} \biggr) \\
    \frac{dx_{art}}{dt} &= Q_{tot} \, \biggl( \frac{x_{lungs}/V_{lungs}}{P_{lungs:p}/B:P} - \frac{x_{art}}{V_{art}} \biggr) \\
    \frac{dx_{ven}}{dt} &= \sum_{t \in \mathcal{T}} \Biggl[ Q_t \, \biggl( \frac{x_t/V_t}{P_{t:p}/B:P} \biggr) \Biggr] - Q_{tot} \cdot \frac{x_{ven}}{V_{ven}} \\
    \end{aligned}
\end{equation}
Subscript $ven$ stands for venous blood. $\mathcal{T}$ represents all the tissues except lungs, arterial and venous blood, small and large intestine, stomach, spleen and pancreas. The difference between lung and Equation \ref{eq_pbpk_t} is that the lungs receive the input from venous blood with a flux equal to $Q_{tot}$, or cardiac output. The arterial blood compartment receives its input from the lungs, while the venous blood compartment receives its input from the outputs of all organs defined in $\mathcal{T}$.

Midazolam (MDZ) is primarily metabolised in the liver by the two enzymes, CYP3A4 and CYP3A5. For MDZ both enzymes catalyse two reactions, leading to the formation of two metabolites,\textit{1-hydroxy} \textit{midazolam} (1-OH-MDZ) and \textit{4-hydroxy} \textit{midazolam} (4-OH-MDZ) \cite{vossen_dynamically_2007,Galetin_midazolam_2004}. For this reason, two mass flows corresponding to MDZ metabolism leave the PBPK system from the liver compartment following intravenous drug administration, as represented in Equation system \ref{eq_liver}.
\begin{equation}\label{eq_liver}
    \begin{aligned}
    \frac{dx_{liv}}{dt} &= Q_{liv} \biggl(\frac{x_{art}}{V_{art}} - \frac{x_{liv}/V_{liv}}{P_{liv:p}/B:P}\biggr) + \sum_{t \in \mathcal{S}} \Biggl[ Q_t \, \biggl( \frac{x_t/V_t}{P_{t:p}/B:P} \biggr) \Biggr] \\
    & \, - MET_{3A4} - MET_{3A5} \\
    MET_{3A4} &= \frac{\tilde{V}_{max,3A4,1-OH} \cdot c_{u,liv}}{K_{M,3A4,1-OH} + c_{u,liv}} + \frac{\tilde{V}_{max,3A4,4-OH} \cdot c_{u,liv}}{K_{M,3A4,4-OH} + c_{u,liv}}\\
    MET_{3A5} &= \frac{\tilde{V}_{max,3A5,1-OH} \cdot c_{u,liv}}{K_{M,3A5,1-OH} + c_{u,liv}} + \frac{\tilde{V}_{max,3A5,4-OH} \cdot c_{u,liv}}{K_{M,3A5,4-OH} + c_{u,liv}}\\
    \end{aligned}
\end{equation}
Subscript $liv$ stands for liver, $\mathcal{S}$ represents the splanchnic organs (spleen, pancreas, stomach, small and large intestine). $c_{u,liv}$ is the unbound liver concentration, equal to $x_{liv} \cdot fu_{t}/V_{liv}$, where $fu_{t}$ is the fraction unbound drug in the tissue. $MET_{3A4}$ and $MET_{3A5}$ are the fluxes representing the reactions catalysed by CYP3A4 and CYP3A5. Subscripts $1-OH$ and $4-OH$ refer to the reactions leading to the formation of 1-OH-MDZ and 4-OH-MDZ. All the chemical reactions are described using \textit{Michaelis-Menten} equations \cite{michaelis_kinetik_1913}, where $\tilde{V}_{max}$ is the \textit{in vivo} maximum reaction rate and $K_M$ is the substrate concentration at which the rate is half of $\tilde{V}_{max}$. $\tilde{V}_{max}$ is function of the \textit{in vivo} enzyme abundance and is derived in equation \ref{eq_Vmax}, as per \cite{rostamihodjegan_simulation_2007}.

\begin{equation}\label{eq_Vmax}
    \tilde{V}_{max} = V_{max} \cdot [CYP] \cdot MPPGL \cdot W_{liv}
\end{equation}
$V_{max}$ is the experimentally determined \textit{in vitro} maximum rate per amount of CYP isoform, in $(pmol/min)/(pmol \, CYP)$. $[CYP]$ is the enzyme amount per amount of microsomal protein\footnote{Vesicles derived from the endoplasmic reticulum abundant in drug metabolising enzymes.} (MP), in $(pmol \, CYP)/(mg \, MP)$. $MPPGL$ is the amount of microsomal protein per gram of liver, in $(mg \, MP)/(g \, liver)$. 
Finally, $W_{liv}$ is the liver weight in grams. 

For simulating the pharmacokinetics in a given population of subjects, the PBPK model parameters, such as organ volumes and blood flows, need to be generated reflecting the population distribution. We developed a simple algorithm for generating the organ volumes and blood flows. Briefly:
\begin{enumerate}
    \item the sex of the subject is extracted;
    \item according to the sex, the mean cardiac output and the parameters for height and body mass index (BMI) distributions are fixed;
    \item height and BMI of the subject are extracted;
    \item the body weight ($BW$, in $kg$) is calculated as $BW = BMI \cdot h^2$, where $h$ is the height in $m$;
    \item the cardiac output ($CO$) is calculated as $CO = \bigl( \frac{h}{h_{mean}} \bigr)^{0.75} \cdot CO_{mean}$, with $h_{mean}$ and $CO_{mean}$ the subjects' mean height and cardiac output, respectively \cite{willmann_development_2007};
    \item organ weights and blood flows were derived by multiplying $BW$ and $CO$ for the respective organs fractions, given in Table \ref{tab_organs_par};
    \item organs volumes were derived by dividing the organ weights with organ densities, reported in Table \ref{tab_organs_comp}.
\end{enumerate}

\clearpage
\section{PBPK parameters}
\begin{table}[ht]
	\captionof{table}{Organs composition}
	\centering
	\begin{tabular}[ht]{ l p{1.3cm} p{2cm} p{1.3cm} p{1.3cm} }
		\toprule
		\textbf{Organs} & {neutral lipids fraction \cite{poulin_prediction_2002}} & {phospholipids fraction \cite{poulin_prediction_2002}} & {water fraction \cite{poulin_prediction_2002}}  & {organ density$^c$} \\
		\midrule
     	Adipose              & 0.79       & 0.002     & 0.18      & 0.916    \\
     	Bone                 & 0.074      & 0.0011    & 0.439     & 1.4303   \\
     	Brain                & 0.051      & 0.0565    & 0.77      & 1.0365   \\
     	Heart                & 0.0115     & 0.0166    & 0.758     & 1.03     \\
     	Muscle               & 0.0238     & 0.0072    & 0.76      & 1.041    \\
     	Skin                 & 0.0284     & 0.0111    & 0.718     & 1.1754   \\
     	Spleen               & 0.0201     & 0.0198    & 0.788     & 1.054    \\
     	Kidney               & 0.0207     & 0.0162    & 0.783     & 1.05     \\
     	Gonads$^a$           & 0.0048     & 0.01      & 0.8       & 1$^e$    \\
     	Lung                 & 0.003      & 0.009     & 0.811     & 1.0515   \\
     	Stomach$^b$          & 0.0487     & 0.0163    & 0.718     & 1.046    \\
     	Small intestine$^b$  & 0.0487     & 0.0163    & 0.718     & 1.046    \\
     	Large intestine$^b$  & 0.0487     & 0.0163    & 0.718     & 1.046    \\
     	Liver                & 0.0348     & 0.0252    & 0.751     & 1.08$^f$ \\
     	Pancreas             & 0.0403$^d$ & 0.009$^d$ & 0.641$^d$ & 1.045    \\
     	Plasma               & 0.0035     & 0.00225   & 0.945     & 1$^e$    \\  
		\bottomrule
		\multicolumn{5}{l}{$^{a}$ Values taken from \textit{Open Systems Pharmacology suite} version 7.1.} \\
		\multicolumn{5}{l}{$^{b}$ Values for stomach, small and large intestine were supposed equal.} \\
		\multicolumn{5}{l}{$^{c}$ Calculated using specific gravity values from \cite{brown_physiological_1997},} \\ 
		\multicolumn{5}{l}{considering that water density is 1 $kg/L$.} \\
		\multicolumn{5}{l}{$^{d}$ values taken from \cite{rodgers_physiologically_2005,rodgers_physiologically_2006}} \\
		\multicolumn{5}{l}{$^{e}$ Gonads and blood density were fixed to 1.} \\
		\multicolumn{5}{l}{$^{f}$ Value taken from \cite{heinemann_standard_1999}.} \\
	\end{tabular}
	\label{tab_organs_comp}
\end{table}


\begin{table}[ht]
	\captionof{table}{Organs weight, blood flows and blood content}
	\centering
	\begin{tabular}[ht]{ l c c c c c c}
		\toprule
		\textbf{Organs} & \multicolumn{2}{c}{weight fraction$^b$} & \multicolumn{2}{c}{blood flow fraction$^a$} & \multicolumn{2}{c}{blood fraction$^c$} \\
		\cmidrule(lr){2-7} 
		& male & female & male & female & male & female \\
		\midrule
		Adipose         & 0.2040  & 0.3220  & 0.0530 & 0.0900 & 0.05   & 0.0850 \\
		Bone            & 0.1620  & 0.1520  & 0.0530 & 0.0500 & 0.07   & 0.07   \\
		Brain           & 0.0210  & 0.0230  & 0.1280 & 0.130  & 0.012  & 0.012  \\
		Heart           & 0.0057  & 0.0055  & 0.0430 & 0.05   & 0.01   & 0.01   \\
		Muscle          & 0.4430  & 0.3380  & 0.1810 & 0.12   & 0.14   & 0.105  \\
		Skin            & 0.0520  & 0.0450  & 0.0530 & 0.05   & 0.03   & 0.03   \\
		Spleen          & 0.0033  & 0.0037  & 0.0320 & 0.03   & 0.014  & 0.0104 \\
		Kidney          & 0.0060  & 0.0067  & 0.2170 & 0.2    & 0.02   & 0.02   \\
		Gonads          & 0.0006  & 0.0002  & 0.0005 & 0.0002 & 0.0004 & 0.0002 \\
		Lung            & 0.0180  & 0.0170  & 1      & 1      & 0.1050 & 0.1050 \\
		Stomach         & 0.0023  & 0.0027  & 0.0110 & 0.01   & 0.01   & 0.01   \\
		Small intestine & 0.0100  & 0.0120  & 0.1060 & 0.12   & 0.038  & 0.038  \\
		Large intestine & 0.0056  & 0.0069  & 0.0430 & 0.05   & 0.022  & 0.022  \\
		Liver           & 0.0320  & 0.0320  & 0.0690 & 0.07   & 0.1    & 0.1    \\
		Pancreas        & 0.0026  & 0.0028  & 0.0110 & 0.01   & 0.006  & 0.006  \\
		Blood           & 0.0767$^d$ & 0.0683$^d$ & - & - & (0,06,0.18)$^e$  & (0.06,0.18)$^e$  \\
		\bottomrule
		\multicolumn{7}{l}{$^{a}$ Organ weight fraction (including blood content) on total body weight \cite{willmann_development_2007}.} \\
		\multicolumn{7}{l}{$^{b}$ Fraction of cardiac output directed to each organ \cite{willmann_development_2007}.} \\
		\multicolumn{7}{l}{$^{c}$ Fraction of blood weight (relative to total blood weight) \cite{valetin_basic_2002}.} \\
		\multicolumn{7}{l}{$^{d}$ Blood fraction on total body weight \cite{valetin_basic_2002}.} \\
		\multicolumn{7}{l}{$^{e}$ (arterial fraction, venous fraction) \cite{valetin_basic_2002}.} \\
	\end{tabular} 
	\label{tab_organs_par}
\end{table}


\begin{table}[ht]
	\captionof{table}{Midazolam related parameters}
	\centering
	\begin{tabular}[ht]{ l c c c}
		\toprule
		\textbf{Parameters} & value & units & references \\
		\midrule
		$B:P$                   & 0.66        &                         & \cite{brill_mdz_pbpk_2016}  \\
		$fu_p$                  & 0.0303      &                         & \cite{brill_mdz_pbpk_2016}  \\
		molecular weight        & 325.77      & $g/mol$                 & \cite{osp_suite}            \\
		$log_{10}P_{ow}$        & 3.13        &                         & \cite{osp_suite}            \\
		$V_{max,3A4,1}$         & 1.96        & $pmol/min/(pmol\,CYP)$  & \cite{Galetin_midazolam_2004}   \\
		$K_{M,3A4,1}$           & 2.69        & $\mu M$                 & \cite{Galetin_midazolam_2004}   \\
		$V_{max,3A4,4}$         & 2.52        & $pmol/min/(pmol\,CYP)$  & \cite{Galetin_midazolam_2004}   \\
		$K_{M,3A4,4}$           & 29          & $\mu M$                 & \cite{Galetin_midazolam_2004}   \\
		$V_{max,3A5,1}$         & 6.7         & $pmol/min/(pmol\,CYP)$  & \cite{Galetin_midazolam_2004}   \\
		$K_{M,3A5,1}$           & 10.7        & $\mu M$                 & \cite{Galetin_midazolam_2004}   \\
		$V_{max,3A5,4}$         & 0.52        & $pmol/min/(pmol\,CYP)$  & \cite{Galetin_midazolam_2004}   \\
		$K_{M,3A5,4}$           & 12.1        & $\mu M$                 & \cite{Galetin_midazolam_2004}   \\
		\bottomrule
	\end{tabular}
	\label{tab_drug_param}
\end{table}

\clearpage
\section{Convergence of the Kucherenko indices}
\begin{figure}[h] 
	\centering
	{\includegraphics[scale=0.35]{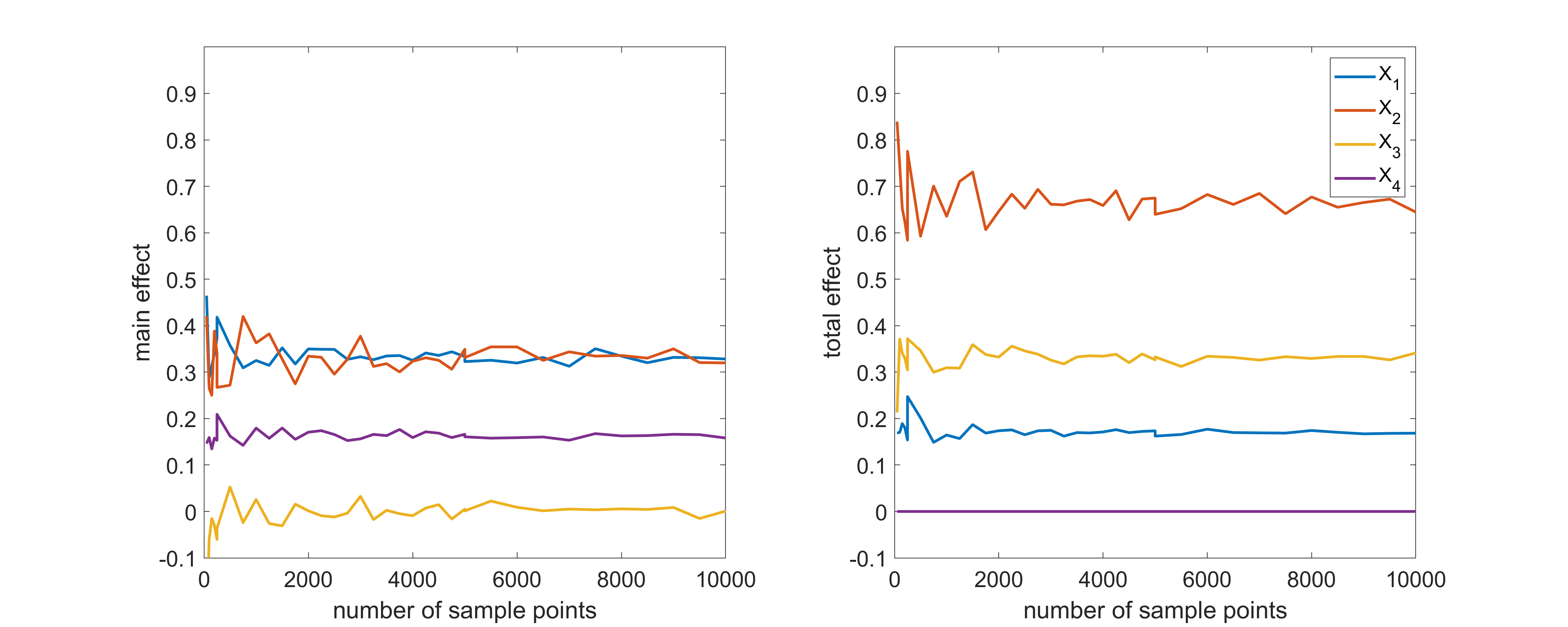}}
	\caption{Convergence plot for the Kucherenko indices of the algebraic model 1, with $\rho=0.7$}
	\label{fig_conv_plot_rho7_mod1}
\end{figure}

\begin{figure}[h] 
	\centering
	{\includegraphics[scale=0.35]{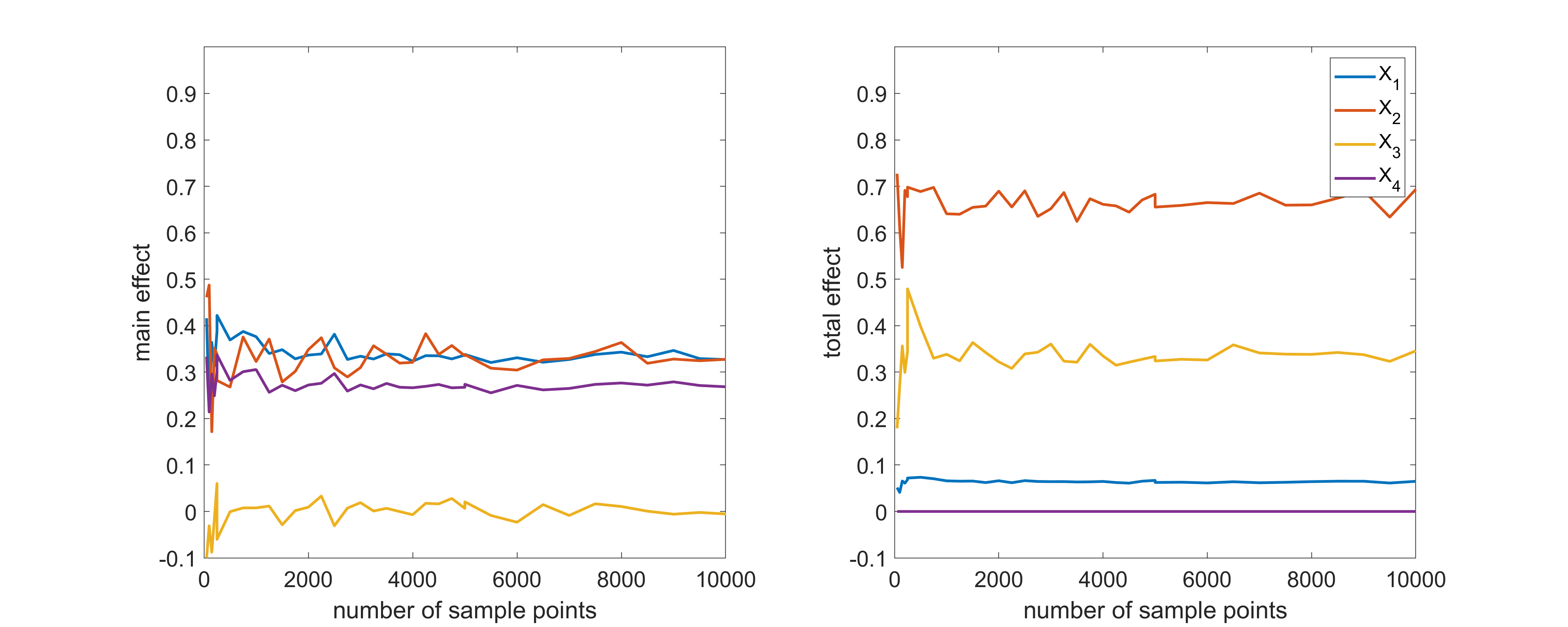}}
	\caption{Convergence plot for the Kucherenko indices of the algebraic model 1, with $\rho=0.9$}
	\label{fig_conv_plot_rho9_mod1}
\end{figure}

\begin{figure}[h] 
	\centering
	{\includegraphics[scale=0.35]{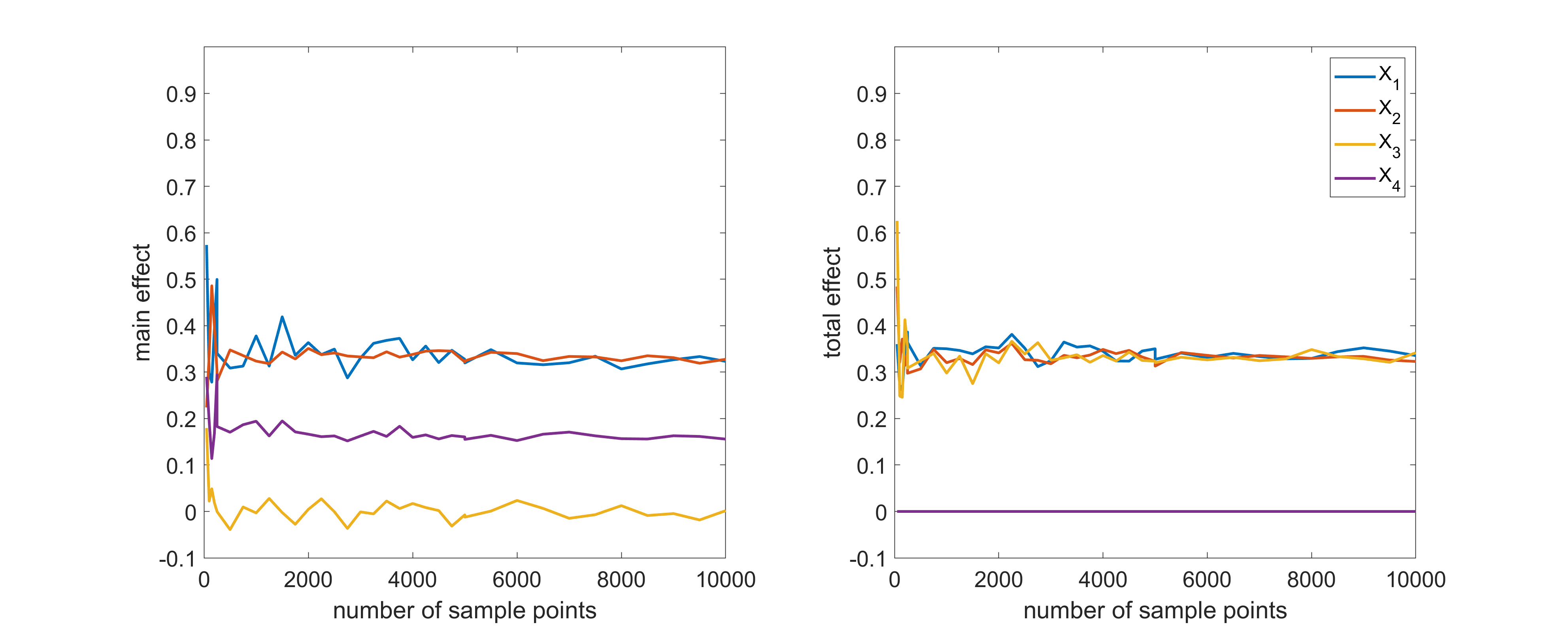}}
	\caption{Convergence plot for the Kucherenko indices of the algebraic model 2, with $\rho=0.7$}
	\label{fig_conv_plot_rho7_mod2}
\end{figure}

\begin{figure}[h] 
	\centering
	{\includegraphics[scale=0.35]{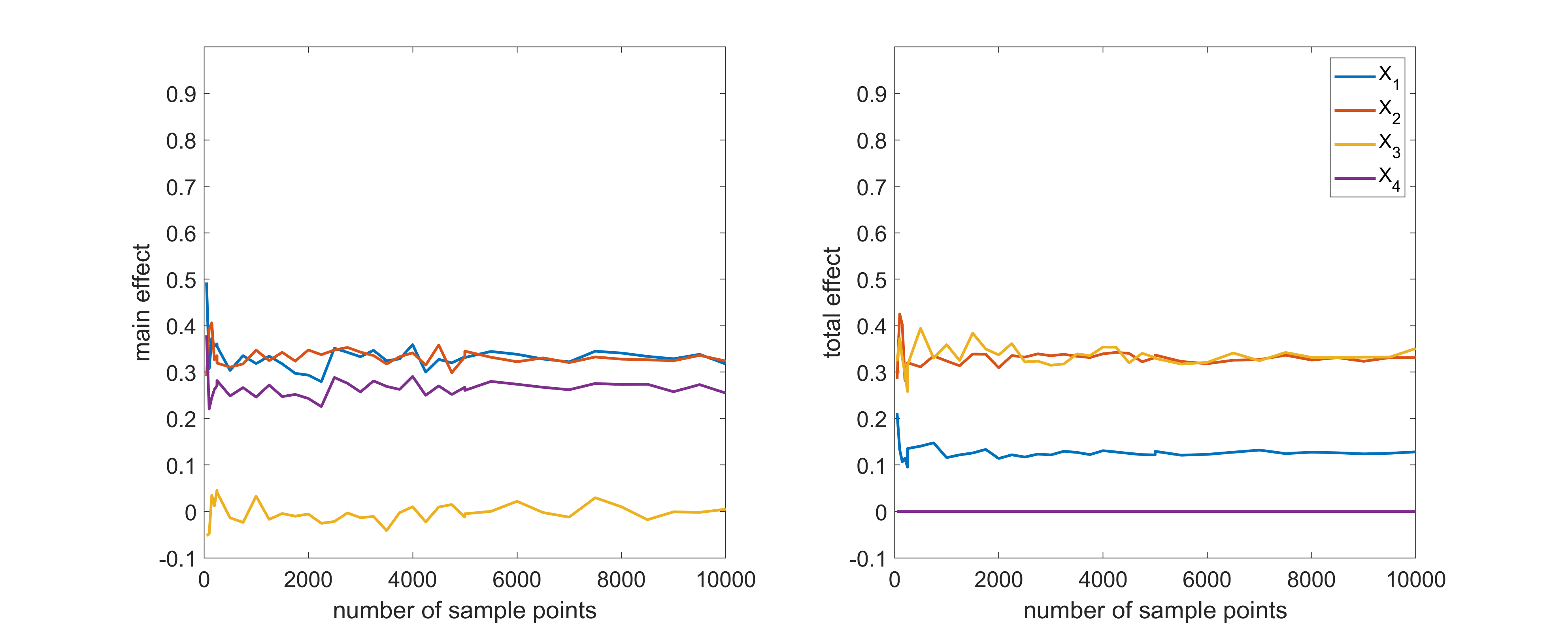}}
	\caption{Convergence plot for the Kucherenko indices of the algebraic model 2, with $\rho=0.9$}
	\label{fig_conv_plot_rho9_mod2}
\end{figure}

\begin{figure}[h] 
	\centering
	{\includegraphics[scale=0.35]{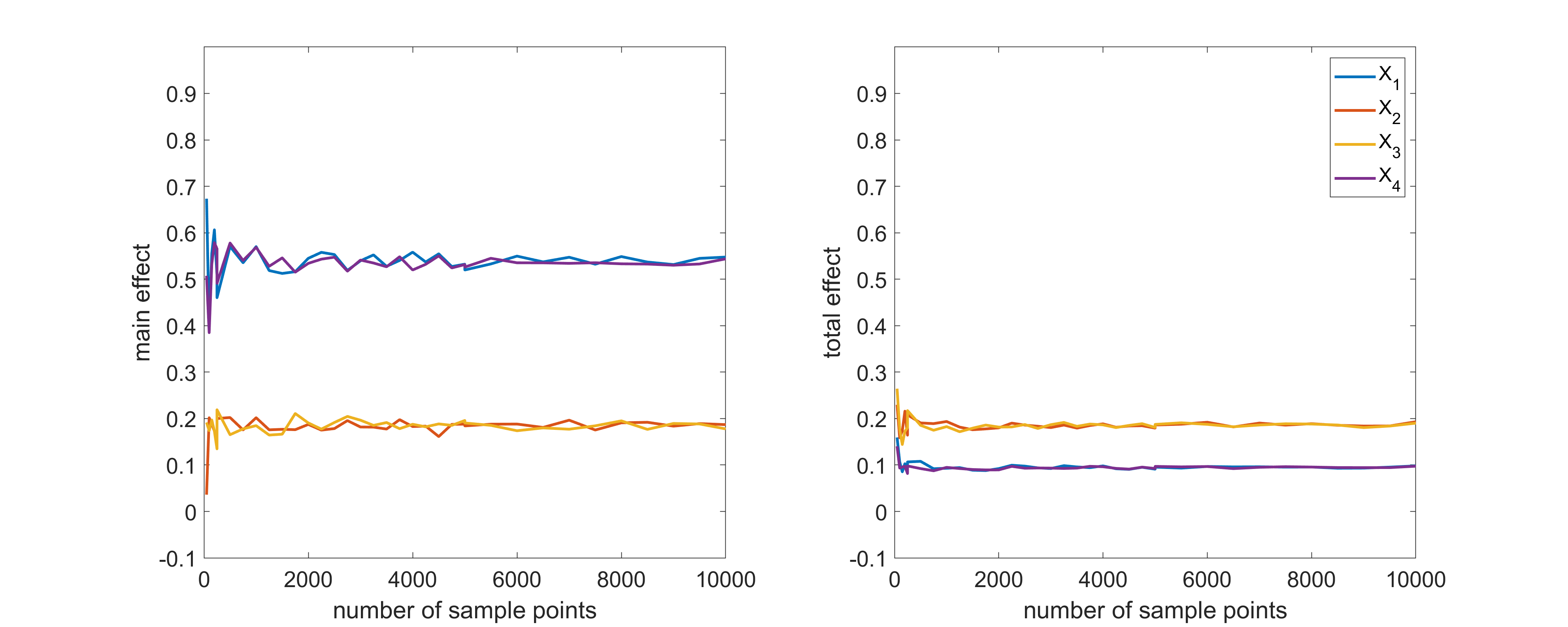}}
	\caption{Convergence plot for the Kucherenko indices of the algebraic model 3, with $\rho=0.7$}
	\label{fig_conv_plot_rho9_mod3}
\end{figure}

\begin{figure}[h] 
	\centering
	{\includegraphics[scale=0.35]{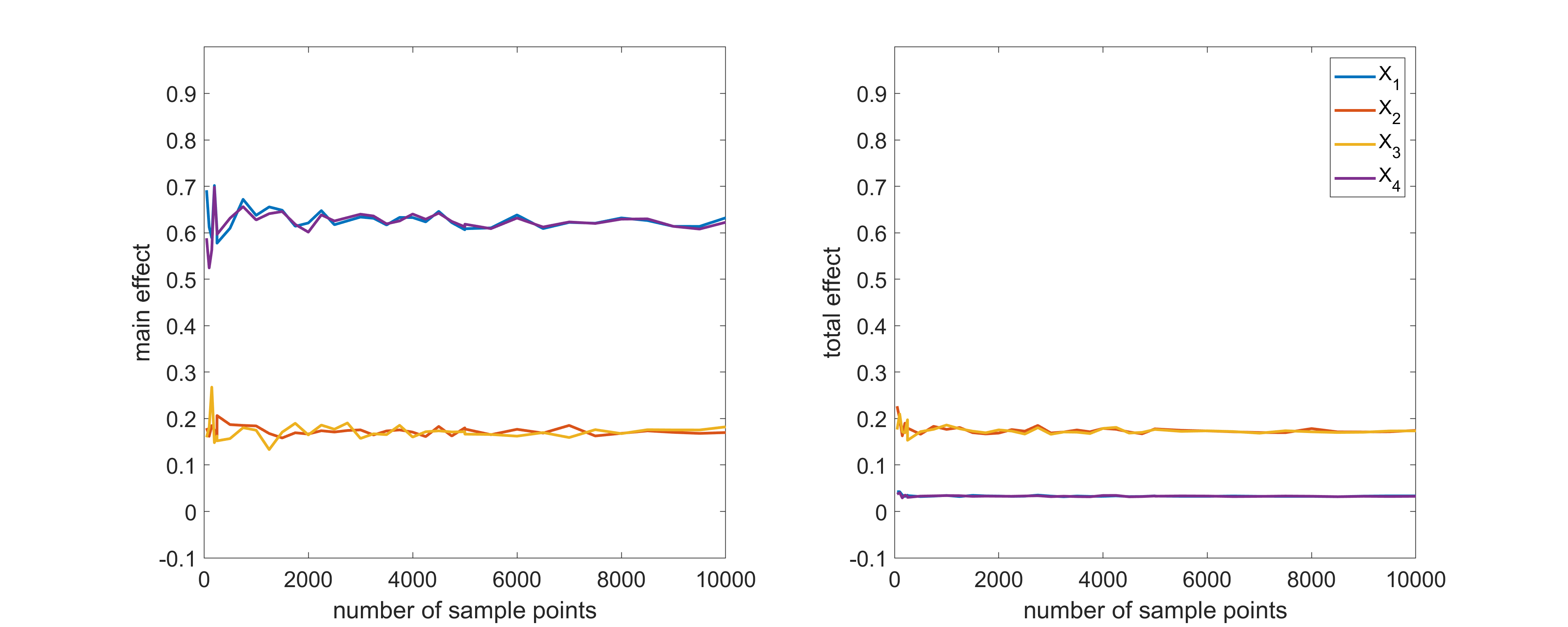}}
	\caption{Convergence plot for the Kucherenko indices of the algebraic model 3, with $\rho=0.9$}
	\label{fig_conv_plot_rho7_mod3}
\end{figure}

\begin{figure}[h] 
	\centering
	{\includegraphics[scale=0.3]{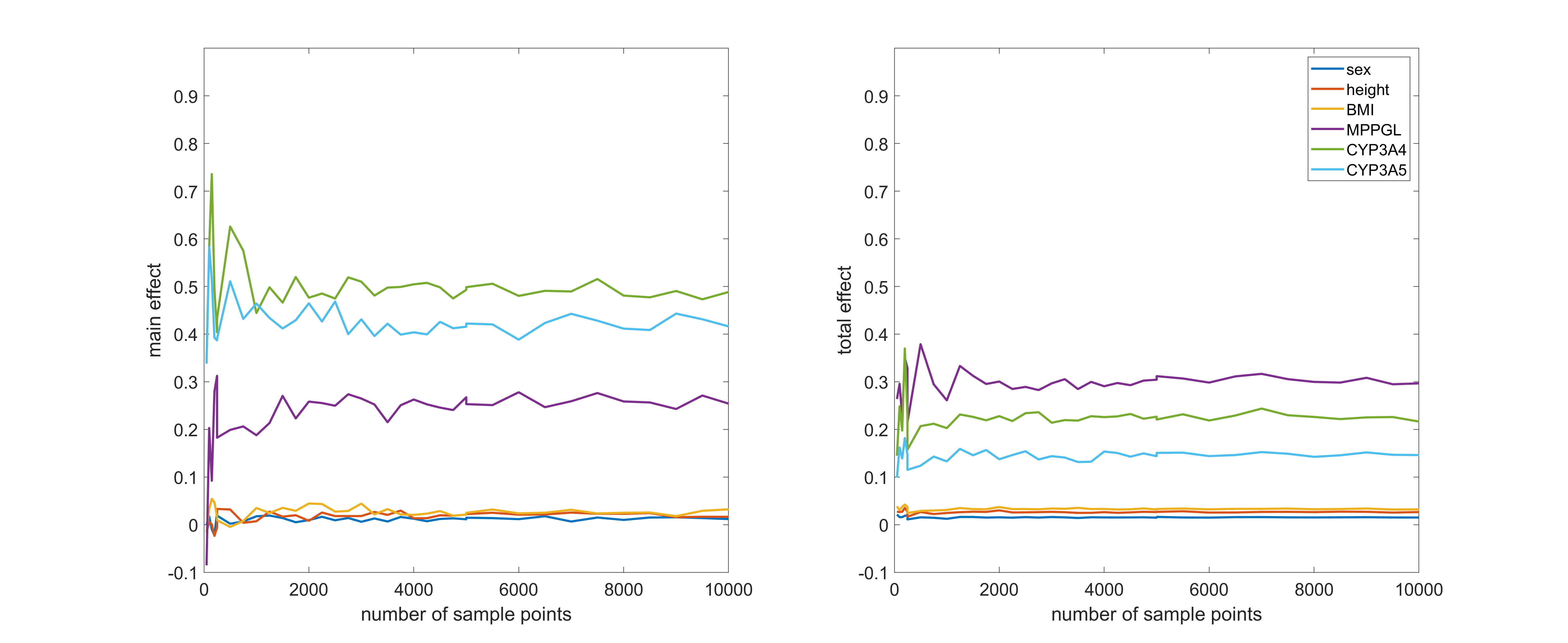}}
	\caption{Convergence plot for the Kucherenko indices of the PBPK model for subjects expressing CYP3A5}
	\label{fig_conv_plot_pbpk}
\end{figure}

Figures \ref{fig_conv_plot_rho7_mod1} to \ref{fig_conv_plot_pbpk} detail the convergence of the Kucherenko indices for the various models that were examined.

\clearpage


\bibliographystyle{unsrt}  

\newpage
\bibliography{Biblio.bib}

\begin{thebibliography}{10}

\bibitem{chmp_ema_guideline_2018}
{CHMP (EMA)}.
\newblock Guideline on the reporting of physiologically based pharmacokinetic
  ({PBPK}) modelling and simulation.
\newblock Technical Report EMA/CHMP/458101/2016, Committee for Medicinal
  Products for Human Use (CHMP), European Medicines Agency (EMA), London, UK,
  December 2018.

\bibitem{fda_pbpkguide_2018}
CDER.
\newblock Physiologically based pharmacokinetic analyses - format and content:
  Guidance for industry, 2018.

\bibitem{melillo_variance_2019}
Nicola Melillo, Leon Aarons, Paolo Magni, and Adam~S. Darwich.
\newblock Variance based global sensitivity analysis of physiologically based
  pharmacokinetic absorption models for bcs i-iv drugs.
\newblock {\em Journal of Pharmacokinetics and Pharmacodynamics}, 46(1):27--42,
  February 2019.

\bibitem{melillo_accounting_2019}
Nicola Melillo, Adam~S. Darwich, Paolo Magni, and Amin Rostami-Hodjegan.
\newblock Accounting for inter-correlation between enzyme abundance: a
  simulation study to assess implications on global sensitivity analysis within
  physiologically-based pharmacokinetics.
\newblock {\em Journal of Pharmacokinetics and Pharmacodynamics},
  46(2):137--154, April 2019.

\bibitem{melillo_inter-compound_2020}
Nicola Melillo, Silvia Grandoni, Nicola Cesari, Giandomenico Brogin, Paola
  Puccini, and Paolo Magni.
\newblock Inter-compound and {Intra}-compound {Global} {Sensitivity} {Analysis}
  of a {Physiological} {Model} for {Pulmonary} {Absorption} of {Inhaled}
  {Compounds}.
\newblock {\em The AAPS Journal}, 22(5):116, August 2020.

\bibitem{mcnally_workflow_2011}
Kevin McNally, Richard Cotton, and George~D. Loizou.
\newblock A {Workflow} for {Global} {Sensitivity} {Analysis} of {PBPK}
  {Models}.
\newblock {\em Frontiers in Pharmacology}, 2:31, 2011.

\bibitem{zhang_sobol_2015}
X-Y Zhang, Mn~Trame, Lj~Lesko, and S~Schmidt.
\newblock Sobol {Sensitivity} {Analysis}: {A} {Tool} to {Guide} the
  {Development} and {Evaluation} of {Systems} {Pharmacology} {Models}.
\newblock {\em CPT: Pharmacometrics \& Systems Pharmacology}, 4(2):69--79,
  February 2015.

\bibitem{daga_physiologically_2018-1}
Pankaj~R. Daga, Michael~B. Bolger, Ian~S. Haworth, Robert~D. Clark, and Eric~J.
  Martin.
\newblock Physiologically {Based} {Pharmacokinetic} {Modeling} in {Lead}
  {Optimization}. 2. {Rational} {Bioavailability} {Design} by {Global}
  {Sensitivity} {Analysis} {To} {Identify} {Properties} {Affecting}
  {Bioavailability}.
\newblock {\em Molecular Pharmaceutics}, 15(3):831--839, March 2018.

\bibitem{yau_gsappkrr_2020}
E.~Yau, A.~Olivares-Morales, M.~Gertz, N.~Parrott, A.~S. Darwich, L.~Aarons,
  and K.~Ogungbenro.
\newblock Global sensitivity analysis of the rodgers and rowland model for
  prediction of tissue: Plasma partitioning coefficients: Assessment of the key
  physiological and physicochemical factors that determine small-molecule
  tissue distribution.
\newblock {\em AAPS J}, 22(2):41, 2020.

\bibitem{liu_investigating_2019}
Dan Liu, Linzhong Li, Amin Rostami-Hodjegan, and Masoud Jamei.
\newblock Investigating {Impacts} of {Model} {Parameters} {Correlations} in
  {Global} {Sensitivity} {Analysis}: {Determining} the most influential
  parameters of a {Minimal} {PBPK} {Model} of {Midazolam}.
\newblock In {\em PAGE 28, Abstr 8875 [www.page-meeting.org/?abstract=8875]},
  Stockholm, Sweden, 2019.

\bibitem{efpia_mid3goodpractices_2016}
EFPIA~MID3 Workgroup, SF~Marshall, R~Burghaus, V~Cosson, SYA Cheung, M~Chenel,
  O~DellaPasqua, N~Frey, B~Hamrén, L~Harnisch, F~Ivanow, T~Kerbusch,
  J~Lippert, PA~Milligan, S~Rohou, A~Staab, JL~Steimer, C~Tornøe, and SAG
  Visser.
\newblock Good practices in model-informed drug discovery and development:
  Practice, application, and documentation.
\newblock {\em CPT: Pharmacometrics \& Systems Pharmacology}, 5(3):93--122,
  2016.

\bibitem{saltelli_what_2013}
Andrea Saltelli, \^Angela Guimaraes~Pereira, Jeroen P. Van~der Sluijs, and
  Silvio Funtowicz.
\newblock What do {I} make of your latinorum? {Sensitivity} auditing of
  mathematical modelling.
\newblock {\em International Journal of Foresight and Innovation Policy},
  9(2/3/4):213, 2013.

\bibitem{aarons_physiologically_2005}
L.~Aarons.
\newblock Physiologically based pharmacokinetic modelling: a sound mechanistic
  basis is needed.
\newblock {\em British Journal of Clinical Pharmacology}, 60(6):581--583, 2005.
\newblock \_eprint:
  https://bpspubs.onlinelibrary.wiley.com/doi/pdf/10.1111/j.1365-2125.2005.02560.x.

\bibitem{bonate_poppkpd_2005}
P.~L. Bonate.
\newblock Recommended reading in population pharmacokinetic pharmacodynamics.
\newblock {\em AAPS J}, 7(2):E363--73, 2005.

\bibitem{jones_basic_2013}
H.~M. Jones and K.~Rowland-Yeo.
\newblock Basic {Concepts} in {Physiologically} {Based} {Pharmacokinetic}
  {Modeling} in {Drug} {Discovery} and {Development}.
\newblock {\em CPT: Pharmacometrics \& Systems Pharmacology}, page~63, 2013.

\bibitem{sasso_pbtk_2010}
A.~F. Sasso, S.~S. Isukapalli, and P.~G. Georgopoulos.
\newblock A generalized physiologically-based toxicokinetic modeling system for
  chemical mixtures containing metals.
\newblock {\em Theor Biol Med Model}, 7:17, 2010.

\bibitem{nestorov_wbpbpk_2007}
I.~Nestorov.
\newblock Whole-body physiologically based pharmacokinetic models.
\newblock {\em Expert Opin Drug Metab Toxicol}, 3(2):235--49, 2007.

\bibitem{jamei_pbpklabel_2016}
Masoud Jamei.
\newblock Recent advances in development and application of
  physiologically-based pharmacokinetic (pbpk) models: a transition from
  academic curiosity to regulatory acceptance.
\newblock {\em Current pharmacology reports}, 2:161--169, 2016.

\bibitem{grimstein_physiologically_2019}
Manuela Grimstein, Yuching Yang, Xinyuan Zhang, Joseph Grillo, Shiew-Mei Huang,
  Issam Zineh, and Yaning Wang.
\newblock Physiologically {Based} {Pharmacokinetic} {Modeling} in {Regulatory}
  {Science}: {An} {Update} {From} the {U}.{S}. {Food} and {Drug}
  {Administration}'s {Office} of {Clinical} {Pharmacology}.
\newblock {\em Journal of Pharmaceutical Sciences}, 108(1):21--25, January
  2019.

\bibitem{tsamandouras_incorporation_2015}
Nikolaos Tsamandouras, Thierry Wendling, Amin Rostami-Hodjegan, Aleksandra
  Galetin, and Leon Aarons.
\newblock Incorporation of stochastic variability in mechanistic population
  pharmacokinetic models: handling the physiological constraints using normal
  transformations.
\newblock {\em Journal of Pharmacokinetics and Pharmacodynamics},
  42(4):349--373, August 2015.

\bibitem{rostamihodjegan_reverse_2018}
Amin Rostami-Hodjegan.
\newblock Reverse {Translation} in {PBPK} and {QSP}: {Going} {Backwards} in
  {Order} to {Go} {Forward} {With} {Confidence}.
\newblock {\em Clinical Pharmacology \& Therapeutics}, 103(2):224--232, 2018.

\bibitem{bai_omicspharmaceutics_2018}
Jane P.~F. Bai, Ioannis~N. Melas, Junguk Hur, and Ellen Guo.
\newblock Advances in omics for informed pharmaceutical research and
  development in the era of systems medicine.
\newblock {\em Expert Opinion on Drug Discovery}, 13(1):1--4, 2018.
\newblock PMID: 29073782.

\bibitem{doki_pbpkcorr_2018}
Kosuke Doki, Adam~S. Darwich, Brahim Achour, Aleksi Tornio, Janne~T. Backman,
  and Amin Rostami-Hodjegan.
\newblock Implications of intercorrelation between hepatic cyp3a4-cyp2c8
  enzymes for the evaluation of drug-drug interactions: a case study with
  repaglinide.
\newblock {\em British journal of clinical pharmacology}, 84(5):972--986, 2018.

\bibitem{garciacremades_mechanistic_2020}
Maria Garcia‐Cremades, Nicola Melillo, Iñaki~F. Troconiz, and Paolo Magni.
\newblock Mechanistic {Multiscale} {Pharmacokinetic} {Model} for the
  {Anticancer} {Drug} 2’,2’-difluorodeoxycytidine ({Gemcitabine}) in
  {Pancreatic} {Cancer}.
\newblock {\em Clinical and Translational Science}, XX, 2020.

\bibitem{rowland_pbpkdrivers_2018}
A.~Rowland, M.~van Dyk, A.~M. Hopkins, R.~Mounzer, T.~M. Polasek,
  A.~Rostami-Hodjegan, and M.~J. Sorich.
\newblock Physiologically based pharmacokinetic modeling to identify
  physiological and molecular characteristics driving variability in drug
  exposure.
\newblock {\em Clin Pharmacol Ther}, 104(6):1219--1228, 2018.

\bibitem{sobol_sensitivity_1993}
Ilya~M. Sobol.
\newblock Sensitivity {Estimates} for {Nonlinear} {Mathematical} {Models}.
\newblock {\em Mathematical modelling and computational experiments},
  1(4):407--414, 1993.

\bibitem{saltelli_making_2002}
Andrea Saltelli.
\newblock Making best use of model evaluations to compute sensitivity indices.
\newblock {\em Computer Physics Communications}, 145(2):280--297, May 2002.

\bibitem{saltelli_global_2008}
Andrea Saltelli, Marco Ratto, Terry Andres, Francesca Campolongo, Jessica
  Cariboni, Debora Gatelli, Michaela Saisana, and Stefano Tarantola.
\newblock {\em Global {Sensitivity} {Analysis}. {The} {Primer}}.
\newblock John Wiley \& Sons, Ltd, January 2008.

\bibitem{oakley_probabilistic_2004}
Jeremy~E. Oakley and Anthony O'Hagan.
\newblock Probabilistic sensitivity analysis of complex models: a {Bayesian}
  approach.
\newblock {\em Journal of the Royal Statistical Society: Series B (Statistical
  Methodology)}, 66(3):751--769, 2004.
\newblock \_eprint:
  https://rss.onlinelibrary.wiley.com/doi/pdf/10.1111/j.1467-9868.2004.05304.x.

\bibitem{iooss_shapley_2019}
Bertrand Iooss and Clementine Prieur.
\newblock Shapley effects for sensitivity analysis with correlatedinputs:
  comparisons with {Sobol} indices, numericalestimation and applications.
\newblock {\em International Journal for Uncertainty Quantification}, 9(5),
  2019.
\newblock Publisher: Begel House Inc.

\bibitem{liu_considerations_2020}
Dan Liu, Linzhong Li, Amin Rostami-Hodjegan, Frederic~Y. Bois, and Masoud
  Jamei.
\newblock Considerations and {Caveats} when {Applying} {Global} {Sensitivity}
  {Analysis} {Methods} to {Physiologically} {Based} {Pharmacokinetic} {Models}.
\newblock {\em The AAPS Journal}, 22(5):93, July 2020.

\bibitem{mara_nonparam_2015}
Thierry~A. Mara, Stefano Tarantola, and Paola Annoni.
\newblock Non-parametric methods for global sensitivity analysis of model
  output with dependent inputs.
\newblock {\em Environmental Modelling \& Software}, 72:173 -- 183, 2015.

\bibitem{do_correlation_2020}
Nhu~Cuong Do and Saman Razavi.
\newblock Correlation {Effects}? {A} {Major} but {Often} {Neglected}
  {Component} in {Sensitivity} and {Uncertainty} {Analysis}.
\newblock {\em Water Resources Research}, 56(3):e2019WR025436, 2020.
\newblock \_eprint:
  https://agupubs.onlinelibrary.wiley.com/doi/pdf/10.1029/2019WR025436.

\bibitem{da_veiga_local_2009}
Sebastien Da~Veiga, Francois Wahl, and Fabrice Gamboa.
\newblock Local {Polynomial} {Estimation} for {Sensitivity} {Analysis} on
  {Models} {With} {Correlated} {Inputs}.
\newblock {\em Technometrics}, 51(4):452--463, November 2009.

\bibitem{li_global_2010}
Genyuan Li, Herschel Rabitz, Paul~E. Yelvington, Oluwayemisi~O. Oluwole, Fred
  Bacon, Charles~E. Kolb, and Jacqueline Schoendorf.
\newblock Global {Sensitivity} {Analysis} for {Systems} with {Independent}
  and/or {Correlated} {Inputs}.
\newblock {\em The Journal of Physical Chemistry A}, 114(19):6022--6032, May
  2010.

\bibitem{xu_extending_2007}
C.~Xu and G.~Gertner.
\newblock Extending a global sensitivity analysis technique to models with
  correlated parameters.
\newblock {\em Computational Statistics \& Data Analysis}, 51(12):5579--5590,
  August 2007.

\bibitem{kucherenko_estimation_2012}
S.~Kucherenko, S.~Tarantola, and P.~Annoni.
\newblock Estimation of global sensitivity indices for models with dependent
  variables.
\newblock {\em Computer Physics Communications}, 183(4):937--946, April 2012.

\bibitem{tarantola_variance-based_2017}
S.~Tarantola and Thierry~A. Mara.
\newblock {Variance}-{based} {sensitivity} {indices} {of} {computer} {models}
  {with} {dependent} {inputs}: {the} {Fourier} {amplitude} {sensitivity}
  {test}.
\newblock {\em International Journal for Uncertainty Quantification}, 7(6),
  2017.
\newblock Publisher: Begel House Inc.

\bibitem{noauthor_simcyp_2019}
CERTARA L.P.
\newblock Simcyp {Simulator} - {Version} 19, 2019.

\bibitem{borgonovo_sensitivity_2016}
Emanuele Borgonovo and Elmar Plischke.
\newblock Sensitivity analysis: {A} review of recent advances.
\newblock {\em European Journal of Operational Research}, 248(3):869--887,
  February 2016.

\bibitem{pianosi_sensitivity_2016}
Francesca Pianosi, Keith Beven, Jim Freer, Jim~W. Hall, Jonathan Rougier,
  David~B. Stephenson, and Thorsten Wagener.
\newblock Sensitivity analysis of environmental models: {A} systematic review
  with practical workflow.
\newblock {\em Environmental Modelling \& Software}, 79:214--232, May 2016.

\bibitem{iooss_review_2015}
Bertrand Iooss and Paul LemaÃ®tre.
\newblock A {Review} on {Global} {Sensitivity} {Analysis} {Methods}.
\newblock In {\em Uncertainty {Management} in {Simulation}-{Optimization} of
  {Complex} {Systems}}, Operations {Research}/{Computer} {Science} {Interfaces}
  {Series}, pages 101--122. Springer, Boston, MA, 2015.

\bibitem{kostewicz_pbpksoftware_2014}
E.~S. Kostewicz, L.~Aarons, M.~Bergstrand, M.~B. Bolger, A.~Galetin, O.~Hatley,
  M.~Jamei, R.~Lloyd, X.~Pepin, A.~Rostami-Hodjegan, E.~Sjogren, C.~Tannergren,
  D.~B. Turner, C.~Wagner, W.~Weitschies, and J.~Dressman.
\newblock Pbpk models for the prediction of in vivo performance of oral dosage
  forms.
\newblock {\em Eur J Pharm Sci}, 57:300--21, 2014.

\bibitem{loehlin_latent_2017}
John~C. Loehlin and A.~Alexander Beaujean.
\newblock {\em Latent {Variable} {Models}: {An} {Introduction} to {Factor},
  {Path}, and {Structural} {Equation} {Analysis}}.
\newblock Routledge, New York, fifth edition edition, 2017.

\bibitem{brown_confirmatory_2015}
Timothy~A. Brown.
\newblock {\em Confirmatory {Factor} {Analysis} for {Applied} {Research}}.
\newblock Guilford Press, New York, second edition edition, January 2015.

\bibitem{Galetin_midazolam_2004}
Aleksandra Galetin, Caroline Brown, David Hallifax, Kiyomi Ito, and J.~Brian
  Houston.
\newblock Utility of recombinant enzyme kinetics in prediction of human
  clearance: Impact of variability, cyp3a5, and cyp2c19 on cyp3a4 probe
  substrates.
\newblock {\em Drug Metabolism and Disposition}, 32(12):1411--1420, 2004.

\bibitem{roy_cyp3a5polymorphism_2005}
J.~N. Roy, J.~Lajoie, L.~S. Zijenah, A.~Barama, C.~Poirier, B.~J. Ward, and
  M.~Roger.
\newblock Cyp3a5 genetic polymorphisms in different ethnic populations.
\newblock {\em Drug Metab Dispos}, 33(7):884--7, 2005.

\bibitem{lolodi_pxrcyp3a_2017}
O.~Lolodi, Y.~M. Wang, W.~C. Wright, and T.~Chen.
\newblock Differential regulation of cyp3a4 and cyp3a5 and its implication in
  drug discovery.
\newblock {\em Curr Drug Metab}, 18(12):1095--1105, 2017.

\bibitem{homma_importance_1996}
Toshimitsu Homma and Andrea Saltelli.
\newblock Importance measures in global sensitivity analysis of nonlinear
  models.
\newblock {\em Reliability Engineering \& System Safety}, 52(1):1--17, April
  1996.

\bibitem{farrell_ave_2009}
Andrew~M. Farrell.
\newblock Insufficient discriminant validity: A comment on bove, pervan,
  beatty, and shiu (2009).
\newblock {\em Journal of Business Research}, 63(3):324 -- 327, 2010.

\bibitem{fornell_evaluating_1981}
Claes Fornell and David~F. Larcker.
\newblock Evaluating {Structural} {Equation} {Models} with {Unobservable}
  {Variables} and {Measurement} {Error}.
\newblock {\em Journal of Marketing Research}, 18(1):39--50, 1981.
\newblock Publisher: American Marketing Association.

\bibitem{berezhkovskiy_valid_2010}
Leonid~M. Berezhkovskiy.
\newblock A {Valid} {Equation} for the {Well}-{Stirred} {Perfusion} {Limited}
  {Physiologically} {Based} {Pharmacokinetic} {Model} that {Consistently}
  {Accounts} for the {Blood}–{Tissue} {Drug} {Distribution} in the {Organ}
  and the {Corresponding} {Valid} {Equation} for the {Steady} {State} {Volume}
  of {Distribution}.
\newblock {\em Journal of Pharmaceutical Sciences}, 99(1):475--485, January
  2010.

\bibitem{vossen_dynamically_2007}
Michaela Vossen, Michael Sevestre, Christoph Niederalt, In-Jin Jang, Stefan
  Willmann, and Andrea~N. Edginton.
\newblock Dynamically simulating the interaction of midazolam and the {CYP3A4}
  inhibitor itraconazole using individual coupled whole-body
  physiologically-based pharmacokinetic ({WB}-{PBPK}) models.
\newblock {\em Theoretical Biology and Medical Modelling}, 4(1):13, March 2007.

\bibitem{michaelis_kinetik_1913}
{Michaelis L.} and {Menten M.L.}
\newblock Die {Kinetik} der {Invertinwirkung}.
\newblock {\em Biochemistry Zeitung}, 49:333 -- 369, 1913.

\bibitem{rostamihodjegan_simulation_2007}
Amin Rostami-Hodjegan and Geoffrey~T. Tucker.
\newblock Simulation and prediction of in vivo drug metabolism in human
  populations from in vitro data.
\newblock {\em Nature Reviews Drug Discovery}, 6(2):140--148, February 2007.
\newblock Number: 2 Publisher: Nature Publishing Group.

\bibitem{rowland_clinical_1995}
Malcolm Rowland and Thomas~N. Tozer.
\newblock {\em Clinical {Pharmacokinetics}: {Concepts} and {Applications}}.
\newblock Lippincott Williams \& Wilkins, third edition, 1995.

\bibitem{cacciari_italian_2006}
E.~Cacciari, S.~Milani, A.~Balsamo, E.~Spada, G.~Bona, L.~Cavallo, F.~Cerutti,
  L.~Gargantini, N.~Greggio, G.~Tonini, and A.~Cicognani.
\newblock Italian cross-sectional growth charts for height, weight and {BMI} (2
  to 20 yr).
\newblock {\em Journal of Endocrinological Investigation}, 29(7):581--593, July
  2006.

\bibitem{WHO:BMI}
World Health~Organization (WHO).
\newblock {\em Body Mass Index - BMI, (accessed September 28, 2020)}, 2020.
\newblock
  \url{https://www.euro.who.int/en/health-topics/disease-prevention/nutrition/a-healthy-lifestyle/body-mass-index-bmi}.

\bibitem{cubitt_sources_2011}
Helen~E. Cubitt, Karen~R. Yeo, Eleanor~M. Howgate, Amin Rostami-Hodjegan, and
  Zoe~E. Barter.
\newblock Sources of interindividual variability in {IVIVE} of clearance: an
  investigation into the prediction of benzodiazepine clearance using a
  mechanistic population-based pharmacokinetic model.
\newblock {\em Xenobiotica}, 41(8):623--638, August 2011.
\newblock Publisher: Taylor \& Francis \_eprint:
  https://doi.org/10.3109/00498254.2011.560294.

\bibitem{noauthor_simcyp_2017}
CERTARA L.P.
\newblock Simcyp {Simulator} - {Version} 17, April 2017.

\bibitem{achour_simultaneous_2014}
Brahim Achour, Matthew~R. Russell, Jill Barber, and Amin Rostami-Hodjegan.
\newblock Simultaneous {Quantification} of the {Abundance} of {Several}
  {Cytochrome} {P}450 and {Uridine} 5'-{Diphospho}-{Glucuronosyltransferase}
  {Enzymes} in {Human} {Liver} {Microsomes} {Using} {Multiplexed} {Targeted}
  {Proteomics}.
\newblock {\em Drug Metabolism and Disposition}, 42(4):500--510, April 2014.

\bibitem{noauthor_matlab_2019}
The {M}athWorks.
\newblock {MATLAB} {R}2019b, the {M}ahworks, inc., {N}atick, {M}assachusetts,
  {U}nited {S}tates, 2019.

\bibitem{Marelli_uqlab_2014}
S.~Marelli and B.~Sudret.
\newblock {UQLab}: a framework for uncertainty quantification in {MATLAB}.
\newblock In {\em Proc. 2nd Int. Conf. on Vulnerability, Risk Analysis and
  Management {(ICVRAM2014)}, Liverpool, United Kingdom}, 2014.

\bibitem{meek_case_2013}
M.~E.~(Bette) Meek, Hugh~A. Barton, Jos~G. Bessems, John~C. Lipscomb, and
  Kannan Krishnan.
\newblock Case study illustrating the {WHO} {IPCS} guidance on characterization
  and application of physiologically based pharmacokinetic models in risk
  assessment.
\newblock {\em Regulatory Toxicology and Pharmacology}, 66(1):116--129, June
  2013.

\bibitem{IPCS_2010}
{International Programme on Chemical Safety (IPCS)}.
\newblock Characterization and {Application} of {Physiologically} {Based}
  {Pharmacokinetic} {Models} in {Risk} {Assessment}.
\newblock Harmonization {Project} {Document}~9, World Health Organization
  (WHO), 2010.

\bibitem{graham_partition_2012}
Helen Graham, Mike Walker, Owen Jones, James Yates, Aleksandra Galetin, and
  Leon Aarons.
\newblock Comparison of in-vivo and in-silico methods used for prediction of
  tissue: plasma partition coefficients in rat.
\newblock {\em Journal of Pharmacy and Pharmacology}, 64(3):383--396, 2012.

\bibitem{berezhkovskiy_volume_2004}
Leonid~M. Berezhkovskiy.
\newblock Volume of {Distribution} at {Steady} {State} for a {Linear}
  {Pharmacokinetic} {System} with {Peripheral} {Elimination}.
\newblock {\em Journal of Pharmaceutical Sciences}, 93(6):1628--1640, June
  2004.

\bibitem{poulin_prediction_2002}
Patrick Poulin and Frank-Peter Theil.
\newblock Prediction of pharmacokinetics prior to in vivo studies. 1.
  {Mechanism}-based prediction of volume of distribution.
\newblock {\em Journal of Pharmaceutical Sciences}, 91(1):129--156, January
  2002.

\bibitem{willmann_development_2007}
Stefan Willmann, Karsten H\"ohn, Andrea Edginton, Michael Sevestre, Juri
  Solodenko, Wolfgang Weiss, J\"org Lippert, and Walter Schmitt.
\newblock Development of a {Physiology}-{Based} {Whole}-{Body} {Population}
  {Model} for {Assessing} the {Influence} of {Individual} {Variability} on the
  {Pharmacokinetics} of {Drugs}.
\newblock {\em Journal of Pharmacokinetics and Pharmacodynamics},
  34(3):401--431, June 2007.

\bibitem{brown_physiological_1997}
Ronald~P. Brown, Michael~D. Delp, Stan~L. Lindstedt, Lorenz~R. Rhomberg, and
  Robert~P. Beliles.
\newblock Physiological {Parameter} {Values} for {Physiologically} {Based}
  {Pharmacokinetic} {Models}.
\newblock {\em Toxicology and Industrial Health}, 13(4):407--484, July 1997.

\bibitem{rodgers_physiologically_2005}
Trudy Rodgers, David Leahy, and Malcolm Rowland.
\newblock Physiologically {Based} {Pharmacokinetic} {Modeling} 1: {Predicting}
  the {Tissue} {Distribution} of {Moderate}-to-{Strong} {Bases}.
\newblock {\em Journal of Pharmaceutical Sciences}, 94(6):1259--1276, June
  2005.

\bibitem{rodgers_physiologically_2006}
Trudy Rodgers and Malcolm Rowland.
\newblock Physiologically based pharmacokinetic modelling 2: {Predicting} the
  tissue distribution of acids, very weak bases, neutrals and zwitterions.
\newblock {\em Journal of Pharmaceutical Sciences}, 95(6):1238--1257, June
  2006.

\bibitem{heinemann_standard_1999}
Axel Heinemann, Friedel Wischhusen, Klaus P\"uschel, and Xavier Rogiers.
\newblock Standard liver volume in the caucasian population.
\newblock {\em Liver Transplantation and Surgery}, 5(5):366--368, 1999.

\bibitem{valetin_basic_2002}
J.~Valetin.
\newblock Basic {Anatomical} and {Physiological} {Data} for {Use} in
  {Radiological} {Protection}: {Reference} {Values}.
\newblock Technical Report~89, International Commission on Radiological
  Protection (ICRP), 2002.

\bibitem{brill_mdz_pbpk_2016}
MJE Brill, PAJ Välitalo, AS~Darwich, B~van Ramshorst, HPA van Dongen,
  A~Rostami–Hodjegan, M~Danhof, and CAJ Knibbe.
\newblock Semiphysiologically based pharmacokinetic model for midazolam and
  cyp3a mediated metabolite 1-oh-midazolam in morbidly obese and weight loss
  surgery patients.
\newblock {\em CPT: Pharmacometrics \& Systems Pharmacology}, 5(1):20--30,
  2016.

\bibitem{osp_suite}
Open Systems~Pharmacology (OSP).
\newblock {OSP} {Suite} - {Version} 7.1,
  (http://www.open-systems-pharmacology.org/), 2017.

\end{thebibliography}

\end{document}